\documentclass[iop]{emulateapj}
\usepackage{natbib,graphicx,epstopdf}
\usepackage{subfigure}

\newcommand{\wspitzer}{\emph{Warm-Spitzer}}
\newcommand{\spitzer}{\emph{Spitzer}}
\newcommand{\kepler}{\emph{Kepler}}

\newcommand{\teff}{\ensuremath{T_{\rm eff}}}

\newcommand{\rearth}{\ensuremath{R_{\oplus}}}

\newcommand{\teq}{\ensuremath{T_{\rm eq}}}

\shorttitle{Low Kepler False-Positive Rate From Warm-Spitzer Observations}
\shortauthors{D{\'e}sert et al.}\def\simgr{\,\hbox{\hbox{$ > $}\kern -0.8em \lower 1.0ex\hbox{$\sim$}}\,}
\def\simle{\,\hbox{\hbox{$ < $}\kern -0.8em \lower 1.0ex\hbox{$\sim$}}\,}

\begin{document}

\title{Low False-Positive Rate of Kepler Candidates Estimated From A Combination Of Spitzer And Follow-Up Observations}

\author{Jean-Michel D\'esert\altaffilmark{1,2},
David Charbonneau\altaffilmark{3}, 
Guillermo Torres\altaffilmark{3},
Fran\c{c}ois Fressin\altaffilmark{3},
Sarah Ballard\altaffilmark{3,4},
Stephen T. Bryson\altaffilmark{5},
Heather A. Knutson\altaffilmark{2},
Natalie M. Batalha\altaffilmark{6},
William J. Borucki\altaffilmark{5},
Timothy M. Brown\altaffilmark{1,7},
Drake Deming\altaffilmark{8},
Eric B. Ford\altaffilmark{9,10},
Jonathan J. Fortney\altaffilmark{11},
Ronald L. Gilliland\altaffilmark{10},
David W. Latham\altaffilmark{3},
Sara Seager\altaffilmark{12}
}

\altaffiltext{1}{CASA, Department of Astrophysical and Planetary Sciences, University of Colorado, 389-UCB, Boulder, CO 80309, USA; desert@colorado.edu}
\altaffiltext{2}{Division of Geological and Planetary Sciences, California Institute of Technology, Pasadena, CA 91125}
\altaffiltext{3}{Harvard-Smithsonian Center for Astrophysics, 60 Garden Street, Cambridge, USA, MA 02138; desert@colorado.edu}
\altaffiltext{4}{University of Washington, Seattle, WA 98195, USA}
\altaffiltext{5}{NASA Ames Research Center, Moffett Field, CA 94035, USA}
\altaffiltext{6}{San Jose State University, San Jose, CA 95192, USA}
\altaffiltext{7}{Las Cumbres Observatory Global Telescope, Goleta, CA 93117}
\altaffiltext{8}{Department of Astronomy, University of Maryland, College Park, MD 20742-2421, USA}
\altaffiltext{9}{University of Florida, Gainesville, FL 32611}
\altaffiltext{10}{Center for Exoplanets and Habitable Worlds, The Pennsylvania State University, University Park, PA 16802}
\altaffiltext{11}{Department of Astronomy and Astrophysics, University of California, Santa Cruz, CA 95064, USA}
\altaffiltext{12}{Massachusetts Institute of Technology, Cambridge, MA 02159, USA}
 
\begin{abstract}

NASA's \kepler\ mission has provided several thousand transiting planet candidates during the four years of its nominal mission, yet only a small subset of these candidates have been confirmed as true planets.
Therefore, the most fundamental question about these candidates is the fraction of {\it bona fide} planets.
Estimating the rate of {\it false positives} of the overall \kepler\ sample is necessary to derive the planet occurrence rate.

We present the results from two large observational campaigns that were conducted with the \spitzer\ Space Telescope during the the \kepler\ mission.
These observations are dedicated to estimating the {\it false positive rate} (FPR) amongst the \kepler\ candidates.
We select a sub-sample of 51 candidates, spanning wide ranges in stellar, orbital and planetary parameter space, and we observe their transits with \spitzer\ at 4.5~\micron.
We use these observations to measures the candidate's transit depths and infrared magnitudes. An authentic planet produces an achromatic transit depth (neglecting the modest effect of limb darkening). Conversely a bandpass-dependent depth alerts us to the potential presence of a blending star that could be the source of the observed eclipse: a false-positive scenario.

For most of the candidates (85\%), the transit depths measured with \kepler\ are consistent with the transit depths measured with \spitzer\ as expected for planetary objects, while we find that the most discrepant measurements are due to the presence of unresolved stars that dilute the photometry.
The \spitzer\ constraints on their own yield FPRs between 5-40\%, depending on the KOIs.
By considering the population of the \kepler\ field stars, and by combining follow-up observations (imaging) when available, we find that the overall FPR of our sample is low.
The measured upper limit on the FPR of our sample is 8.8\% at a confidence level of 3$\sigma$.
This observational result, which uses the achromatic property of planetary transit signals that is not investigated by the \kepler\ observations, provides an independent indication that \kepler's false positive rate is low.

\end{abstract}



\section{Introduction}
\label{intro}

Launched in March of 2009, NASA's \kepler\ mission is a space-based photometric telescope designed to address important questions on the frequency and characteristics of planetary systems around Sun-like stars, and to search for transiting Earth-analogs~\citep{Borucki10a}.
Statistical answers to these questions are required in order to constrain planetary formation and evolution scenarios.
\kepler\ detects transiting planetary candidates signals through continuous photometric monitoring of about 160\,000 stars at high photometric precision (e.g.,~\citealt{Borucki10a,Borucki10b,Koch10a,Gilliland10a}).
This unprecedented sample of potential exoplanets has become an immense resource for statistical studies of the properties and distributions of planets around main-sequence stars (e.g.,\citealt{Youdin11,Tremaine12, Wu12}). This ensemble of candidates is also necessary for determining the occurrence rate of exoplanets (e.g., \citealt{Howard12,Fressin13}), and more specifically of Earth-size planets in the habitable zone of their parent stars (e.g., \citealt{Catanzarite11,Traub12,Petigura13,Dressing13}).

The mission has led to the detection of 2740 {\it planetary candidates} during the first two years of operation~\citep{Batalha12,Burke14}.
However, only a small subset of these candidates have been confirmed as true planets.
This is because asserting the planetary nature of a transit signal requires significant observational follow-up and computational efforts that are unachievable in a practical sense for every detected candidate.
We do not expect all the signals to be due to planets: many astrophysical phenomena can reproduce a similar lightcurve to that of a transiting
planet~\citep{Brown03}.
Indeed, false-positive contamination is one of the main challenges facing transit surveys such as \kepler.
During the last decade, ground-based surveys dedicated to the search of transiting planets have spent considerable effort in confirming the planetary nature of photometrically detected candidates~\citep{Alonso04,Bakos07,Collier07,Moutou09}.
These surveys have established that false positives usually outnumber true planetary systems by a large factor.
It has been shown that from 80 to 90\% of the candidates are false-positives for the most successful ground-based exoplanet surveys (e.g., \citealt{Latham09}). 
For these reasons, the true false positive rate of \kepler\ remains an active research area because false positives can critically bias estimates of planet occurrence rates (e.g.,~\citealt{Morton11,Morton12,Fressin13}). 
This is the subject of the current paper.

The \kepler\ survey poses new challenges for dynamically confirming (using radial velocity or transit timing variation) the planetary nature of candidates. 
This is because of intrinsic characteristics of the \kepler\ target sample such as the large number of candidates, the candidates' small size (presumably of low mass), and the faintness of the host stars.
Consequently, we must develop new methods to determine the origin of \kepler\ detectable signals.  One method consists of in-depth
statistical validation of candidates by ruling out false positive scenarios one-by-one (e.g., BLENDER,~\citealt{Torres04,Torres11,Fressin11}); it fully exploits the information from the shape of a transit lightcurve~\citep{Seager03}.
The goal of this method is to demonstrate {\it statistically} that a transit signal is more likely to be of planetary origin than to be a false positive. In the case of \kepler, this is generally made possible using follow-up observations, such as spectroscopy, imaging, and multi-wavelength transit photometry (including
with \spitzer).
This was demonstrated in the case of the first validation of a Super-Earth~\citep{Torres11}.
However, each candidate validated by this method requires intense observational and computational follow-up work.
In particular, the follow-up strategies adopted by the \kepler\ team are summarized in~\cite{Batalha10a}, and often require substantial efforts and resources. 
Therefore, it is impractical at present to apply the BLENDER method to each individual \kepler\ transit signal.
Yet at the same time, we require the fractional values of {\it bona fide} planets, or of astrophysical false
positives, to accurately determine the occurrence of planetary systems from \kepler\ ~\citep{Fressin13}.

There are currently several approaches to estimate the {\it False Positive Rate} (FPR) of the \kepler\ sample. 
\cite{Coughlin14} studied the effect of contamination on the FPR due to the design of \kepler\ itself, such as direct PRF (pixel response function), antipodal reflections, CCD cross-talks, or columns anomalies.
The contamination sources are eclipsing binaries, variable stars, and other transiting planets and results in a significant number of the known KOIs to be false positives.
\cite{Coughlin14} performed period-ephemeris matching among all transiting planet, eclipsing binary, and variable star sources. They examined the full KOI list and found that 12\% of KOIs are false-positives due to contamination.
Other approaches use generic arguments about the \kepler\ signals to infer the overall \kepler\ FPR (e.g.,~\citealt{Morton11,Morton12,Fressin13}).
There are also parallel attempts to estimate the FPR of targeted specific samples of Kepler Objects of Interest (KOIs).
For example, studies have focused on close-in gas giant planet candidates (e.g.,~\citealt{Santerne12,Colon12}), or on the multiple planet system candidates~\citep{Rowe14}. 
The latter sample contains less than a percent of false positives~\citep{Latham11,Lissauer12,Lissauer14}. 
Other methods use a proxy of the host stars' mean density to estimate the \kepler\ FPR  (e.g.,~\citealt{Sliski14}).

In this paper, we conduct two campaigns to measure transit depths of KOIs with \spitzer, and combine these observations to followup studies, in order to assess the overall FPR of these samples.
We adopt an approach that expands significantly the number of KOIs that are examined using multi-wavelength photometry.
Our project focuses primarily on smaller size candidates, such as mini-Neptune and Super-Earth size objects, compared to previous targeted sample studies.
We select a sample of 51 candidates, measure their transit depth at 4.5 microns with IRAC, and combine these observations with complementary follow-up studies and information from \kepler\ in order to derive the false positive probability (FPP) for each object.
Our method is based on the fact that the relative depth of a planetary transit is achromatic (neglecting the modest effect of limb-darkening), but not for a blend. 
In contrast, a blend containing a false positive, for instance an eclipsing binary, can yield a depth that can vary significantly with the instrument bandpass and stellar temperatures.
The amplitude of this effect increases correspondingly as the difference in wavelength between the two bandpasses increases.
Since \kepler\ observes through a broad bandpass at visible wavelengths, large color-dependent effects for false positives caused by the presence of blended cool stars can be revealed at infrared wavelengths.
We first applied this method by combining \spitzer\ data from this program and \kepler\ data in~\cite{Fressin11}.

The two Science Exploration \spitzer\ programs, which form the core of the data presented in the current paper, have been an active part of the attempts to validate KOIs.
About 20\% of the total amount of \spitzer\ telescope time allocated for this project has already been used in publications dedicated to the confirmation or the validation of 22 Kepler planets.
These are Kepler-10c \citep{Fressin11}, Kepler-11b \citep{Lissauer11}, Kepler-14b \citep{Buchhave11}, Kepler-18b,c \citep{Cochran11}, Kepler-19b \cite{Ballard11}, Kepler-20b,c \cite{Gautier12}, Kepler-22b \citep{Borucki12}, Kepler-25b,c \citep{Steffen12}, Kepler-26c \citep{Steffen12}, Kepler-32b \citep{Fabrycky12}, Kepler-37b \citep{Barclay13}, Kepler-49b,c \citep{Steffen13}, Kepler-61b \citep{Ballard13}, Kepler-62e \citep{Borucki13}, Kepler-68b \citep{Gilliland13}, Kepler-410A b \citep{Eylen13}, Kepler-93b \citep{Ballard14} (see Table~\ref{tab:fpp} for the correspondance between \kepler\ names and KOI numbers).
Furthermore, some of the KOIs of the current study are already confirmed or validated as planets without using the \spitzer\ data.
In particular, \citet{Rowe14} validate 851 planets in multiple-planet system candidates (including 11 KOIs used in our study) by applying statistical arguments \citep{Latham11, Lissauer12, Lissauer14} on the Q1-Q8 \kepler\ data.
While the radial velocity technique allowed the confirmation of the planetary nature of Kepler-89d \citep{Weiss13}, Kepler-94b \citep{Marcy14}, Kepler-102d,e \citep{Marcy14}.
Nevertheless, the present work disregard previous validation or confirmation of individual object in order to treat the whole KOI list followed with \spitzer\ as a statistical ensemble; this is necessary to estimate the FPR of this sample.
Finally, 150 hours (11\%) of time from these two Exploration Science Programs were used to study the atmospheres of Kepler-detected hot Jupiters detected by monitoring their secondary eclipses~\citep{Desert11b,Fortney11,Desert11c}.

This paper is organized as follows: we first describe the different types of astrophysical false positives that we are concerned with
(Section~\ref{sec:fptype}). We then present the sample of candidates that were selected to conduct this study (Section~\ref{sec:select}). 
The \spitzer\ observations and results are presented in Sections~\ref{sec:spfollowup} and~\ref{sec:spconstraint}. 
The combination of various observational constraints (Section~\ref{sec:observations}) allows us to estimate the FPR of our \kepler\ sample (Section~\ref{sec:FPP}). 
We finally discuss the implications of our findings in Section~\ref{sec:discussion}, in particular, in the context of other studies.


\section{Astrophysical False Positives In The Kepler Signals}
\label{sec:fptype}

There are a variety of astrophysical phenomena that can mimic the signal of a transiting planet passing in front of a main sequence star targeted by \kepler. 
These events are produced by additional stars falling within the same aperture as the target star (presumed to be the brighter star), and significantly diluting the total light observed by Kepler.
More specifically, the kinds of false positives that we are concerned with in our study include background or foreground eclipsing binaries (EBs), blended within the Kepler aperture of the target star, as well as those that are physically associated to the target star. 
We refer to these latter configurations as hierarchical triples (HTs).
HTs often cannot be resolved in high-angular resolution imaging.
In this paper, we check for the presence of a stellar companion by looking at how transit depths vary between the \kepler\ and \spitzer\ bandpasses.
The dilution by a stellar companion, blended in the \kepler\ aperture of a planet host star of interest, can be responsible for variations in the wavelength-dependent transit depths measured for a planet.
We do not consider the case where the contaminating star is itself transited by a planet as a potential false positive scenario.

Before searching for false-positives in \kepler, it is important to recall the major vetting steps that each \kepler\ target goes through.
First, a comprehensive study was applied when assembling the Kepler Input Catalog (KIC;~\citealt{Latham05,Batalha10a,Batalha10b,Brown11}), leading to the identification of some EBs and stellar giants, hence avoiding their continuous monitoring with \kepler.
About 160\,000 stars were carefully selected from the KIC catalog and were continuously monitored photometrically with \kepler\ (\citealt{Jenkins10a}).
\citet{Batalha10b} explains the detection of transit events and the vetting processes that are then applied to reject the most common false positive scenarios.
Transit-like signals are identifiable from TCEs (Threshold-Crossing Events) using the \kepler\ photometry alone. 
The \kepler\ team adopted a detection threshold of $7.1~\sigma$ for the transit so that no more than one spurious signal can occur from purely random fluctuations amongst the 160\,000 stars.
In practice, the process of vetting from TCEs to KOIs involves several qualitative steps that could affect the \kepler\ FPR. 
\citet{Christiansen13} have validated the integrity of this threshold while \citet{Coughlin14} report that TCERT (Threshold Crossing Event Review Team) is 92.9\% effective in detecting false-positives for KOIs from Q1-Q8.

The \kepler\ pipeline identifies grazing EBs by searching for even/odd transit depth differences or by looking for the presence of a clear signature of secondary eclipses.
Giant star-eclipsed-by-a-dwarf star scenarios are detected by recognizing that the primary star is a giant, thereby implying that the size of the transiting body must itself be stellar (e.g.,~\citealt{Gilliland10b,Huber13}).
The detection of the shift in the photocenter at a significance level greater than 3$\sigma$ and the comparison of the difference of in- and out-of transit images show the true source location.
Interestingly, this technique permits the identification of potential contamination by unresolved close-by EBs in an efficient manner~\citep{Jenkins10b,Bryson13}.
However, even for high transit SNR candidates, some blended binary scenarios remain undetectable through the vetting processes.
Therefore, estimating the FPR from \kepler\ requires knowledge of the probability of encountering such blend scenarios.  

Throughout this paper, we follow the notation introduced by~\cite{Torres11}: the objects that comprise a blended binary system are
referred to as the `secondary' and `tertiary', and the candidate star
host is referred to as the `primary'. The distance along the
line-of-sight between the binary system and the main star is
parametrized in terms of the difference in distance modulus, $\mu$. 
The notation applies to every astrophysical false positive scenario.


\section{Selection of the Kepler Object of Interests}
\label{sec:select}

\begin{figure*}[h!]
\begin{center}
\includegraphics[width=7in]{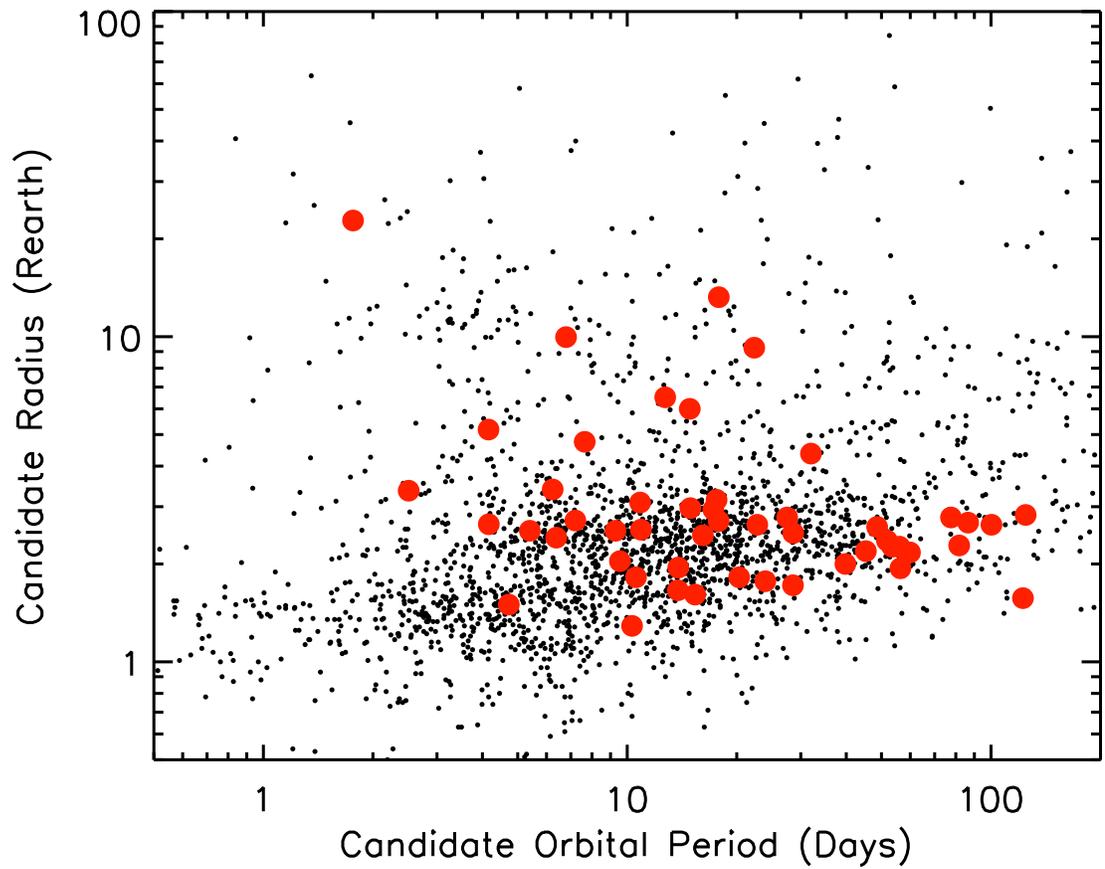}
\caption{Candidate radii as function of their orbital periods (black point) for all the {\it Kepler Object of Interests} (KOIs) presented in \cite{Batalha12}. Overplotted in red are the 51 KOIs that we targeted to estimate the {\it false positive rate} from \kepler\ and that are presented in this paper. We observe these 51 objects during transit in the near-Infrared with \spitzer. Our selected sample spans a wide range of periods and sizes.}
\label{fig:rpper}
\end{center}
\end{figure*}

The first \spitzer\ follow-up program comprised 36 of the first 400 KOIs identified by the Kepler survey~\citep{Borucki11}.
A second set of 23 candidates was selected from the 2335 KOIs compiled by~\cite{Batalha12} for the second \spitzer\ program.
We present the observations and the results for the 36 KOIs from the first program and 15 KOIs from the second program; the current project uses in total 51 KOIs. 
The remaining 8 targets from the second program were not observed yet at the time of the present analysis (KOIs-248.03, 1686.01, 2290.01, 2124.01, 2311.01, 2418.01, 2474.01, 2650.01).
KOIs-248.03 and 2650.01 have recently been validated as Kepler-49d, and Kepler-395c, respectively~\citep{Steffen13, Rowe14}. Importantly, we select these two ensembles with very different criteria.
For the first set, our goal was to derive the FPR of a sub-sample by following-up candidates that represent the diversity of KOIs initially found.
To reach this goal, the first sample is chosen to cover representative ranges of orbital periods, transit depths, stellar types and magnitudes that the first 400 KOIs could allow.
The size of this sample includes about 10\% of the known KOIs at the time.
The ranges of the second set are more tightly constrained: we select candidates for which the expected planetary radius $R_{p}$ would be less than 1.6 \rearth\ and the $\teq\ < 350~K$ (using a stellar temperature \teff\ estimated from the KIC).
Overall, our sample of stellar hosts span a range of KIC estimated temperature from $\teff=3700~K$ up to $\teff=9000~K$.
For both ensembles, the requirement was imposed to accept targets with predicted transit depths detection of at least $3~\sigma$ (scaled from the \kepler\ value), achievable with three or less transits observed with \spitzer.
The selected candidate radii as function of their periods are presented in Figure~\ref{fig:rpper}.


\section{Follow-up observations of selected \kepler\ candidates with \spitzer\ }
\label{sec:spfollowup} 

\subsection{\spitzer\ observations}
\label{sec:spitzerobs}

We use \wspitzer/IRAC~\citep{Werner04,Fazio04} at 4.5~\micron\ to observe transits of the 51 selected KOIs between May 2010 and July 2012.
We obtained these observations as part of two large Science Exploration Programs (program ID 60028 and 80117).
In total, 1400~hours of \spitzer\ time is used for the follow-up of \kepler\ targets.
Of this time, 800~hours of observations are used to complete the first program (60028) and 600~hours are dedicated to the second program (80117).
150~hours of this time were used to study the atmospheres of hot-Jupiters detected by \kepler\ during secondary eclipses~\citep{Desert11b,Fortney11,Desert11c}.
The remaining 1250~hours are dedicated to validating KOIs and estimating the FPR in the \kepler\ data, and are the focus of this paper.
A total of 157 \spitzer\ AORs (Astronomical Observation Requests) have been submitted for these two programs.
This paper focuses on the study of the 51 KOIs presented in Section~\ref{sec:select} that have been observed with \spitzer\ during 95 visits (AORs).

For most of the targeted stars, the data were obtained in a continuous staring full array mode ($256\times256$ pixels) with exposure times of 12 or 30~s, depending on the brightness of the star of interest.
We used the subarray mode of IRAC for the brightest host stars. In this mode, only a $32\times32$-pixel part of the detector is used; this covers a $38\times38\;\mathrm{arcsec}^2$ field of view (pixel size of 1.2\arcsec) and allows for higher cadences (0.2~s exposures).
We choose to put our target at the default pointing position in the center of the field-of-view in order to avoid known hot pixels and bad columns.
This area of the detector is well characterized since it has been extensively used for extrasolar planet studies.
An offset is applied to a few KOIs in order to avoid the presence of a close-by bright target on the same line or column.
The ephemerides of the KOIs were taken from the KFOP database, which are now available on the CFOP website\footnote{https://cfop.ipac.caltech.edu/home/} and we ensured that each visit lasted approximately 2.5 times the transit duration.
We observe 29 KOIs, amongst the 51 presented here, during multiple transit events in order to improve the SNR on the combined lightcurves.
Tables~\ref{tab:spitzer1} and~\ref{tab:spitzer2} list these observations for each program respectively.

\subsection{\spitzer\ photometry}
\label{sec:photometry}

We use the BCD files (Basic Calibrated Data) produced by the \spitzer/IRAC pipeline.
These files include corrections for dark current, flat fielding, pixel non-linearity, and conversion to flux units.
\cite{Desert09} describes the method used to produce photometric time series in each channel from the BCD files.
The method consists of finding the centroid position of the stellar point-spread-function (PSF) and performing aperture photometry.
We first convert the pixel intensities to electrons based on the detector gain and exposure time provided in the FITS headers.
This facilitates the evaluation of the photometric errors.
We extract the BJD date for each image from the FITS headers and compute it to mid-exposure.
We then correct for transient pixels in each individual image using a 20-point sliding median filter of the pixel intensity versus time.
To do so, we compare each pixel's intensity to the median of the 10 preceding and 10 following exposures at the same pixel position and we replace outliers greater than $4~\sigma$ with their median value.
The fraction of pixels that we replace varies between 0.15 to 0.5\%.

The centroid position of the stellar PSF is then determined using DAOPHOT-type Photometry Procedures, \texttt{GCNTRD}, from the IDL Astronomy Library\footnote{{\tt http://idlastro.gsfc.nasa.gov/homepage.html}}.
We use the \texttt{APER} routine to perform aperture photometry with a circular aperture of variable radius.
For each visit, we search for the best aperture size ranging between 1 and 8 pixels radii in steps of 0.5 pixel.
We propagate the uncertainties as a function of the aperture radius and we adopt the size that provides the smallest errors.
We notice that the SNR does not vary significantly with the aperture radii for all the dataset.
The final aperture sizes are set between $2.5$ and $3.5$~pixels depending on the KOIs.

We determine the background level for each frame from two methods.
The first method uses a fit of a Gaussian to the central region of a histogram of counts from the full array, where the background values are defined by the peak position of this Gaussian.
The second method uses the measure of the median value of the pixels inside an annulus centered around the star, with inner and outer radii of $12$ and $20$ pixels respectively, to estimate the background overall level.
Both estimates produce similar results.
The contribution of the background to the total flux from the stars is low for all observations, from 0.1\% to 1.0\% depending on the images, and fairly constant for each AOR.
We find that the residuals from the final light curve modeling are minimized by adopting the center of the Gaussian fits.
After producing the photometric time-series, we use a sliding median filter to select and trim outliers greater than $5\sigma$, which correspond to less than two \% of the data.  We also discard the first half-hour of all observations, which is affected by a significant telescope jitter before stabilization.

Six AORs, corresponding to a total of 30~hours, were gathered at levels above 30,000 DN (in the raw data); this is a level where the detector response tends to be non-linear by several percent.
In order to avoid misinterpreting these data, we do not consider AORs that are above the range of linearization correction. 
Therefore, these six AORs are not used in this work and they are not presented in Table~\ref{tab:spitzer1} and~\ref{tab:spitzer2}.

\subsection{Determination of the transit depths from Spitzer lightcurves}
\label{sec:spdepth}

As described in \cite{Desert11a}, we use a transit light curve model multiplied by instrumental decorrelation functions to measure the transit parameters and their uncertainties from the \spitzer\ data.
We compute the transit light curves with the IDL transit routine \texttt{OCCULTSMALL} from \cite{Mandel02}.
This model depends on the following parameters: the planet-to-star radius ratio $R_p / R_\star$, the orbital semi-major axis to stellar-radius ratio (system scale) $a / R_\star$, the impact parameter $b$, the time of mid transit $T_c$, and limb darkening coefficients.

The measured parameter of interest here is the transit depth.
We fix $T_c$, $a / R_\star$ and $b$ to their values measured from the \kepler\ photometry.
The SNR of our observations is low compared to typical \spitzer\ observations of brighter transiting planets.
The limb darkening effect is negligible at this level of precision; the coefficients are set to zero.
We assume an eccentricity of zero for the KOIs orbit since this parameter does not affect the transit depth measurements at the level of precision we are working with.
Only $R_p / R_\star$ is set as a free parameter to represent the astrophysical signal.

The \spitzer/IRAC photometry is known to be systematically affected by the so-called pixel-phase effect (see e.g.,~\citealt{Charbonneau05}).
This effect is seen as oscillations in the measured fluxes with a period of approximately 70~minutes (period of the telescope pointing jitter) for data secured prior October 2010, and 40 minutes for data secured after.
By October 2010 the \spitzer\ engineering team was able to correlate the pointing wobble with the cycling of a heater used to keep a battery within its operating temperature range. Following this discovery, the \spitzer\ team significantly reduced the amplitude and the period of the pointing wobble.
The amplitude of this effect varies between $1$ and $2\%$, peak to peak, depending on the position of the star in the array.
We decorrelated our signal in each channel using a linear function of time for the baseline (two parameters) and three types of functions to correct the data for the intrapixel variations: a linear function of the PSF position (two parameters), a quadratic function (four parameters) and a quadratic with a cross term (five parameters).
\citet{Desert09} describes in detail the last function.

\begin{figure*}
\begin{center}
 \includegraphics[width=6.5in]{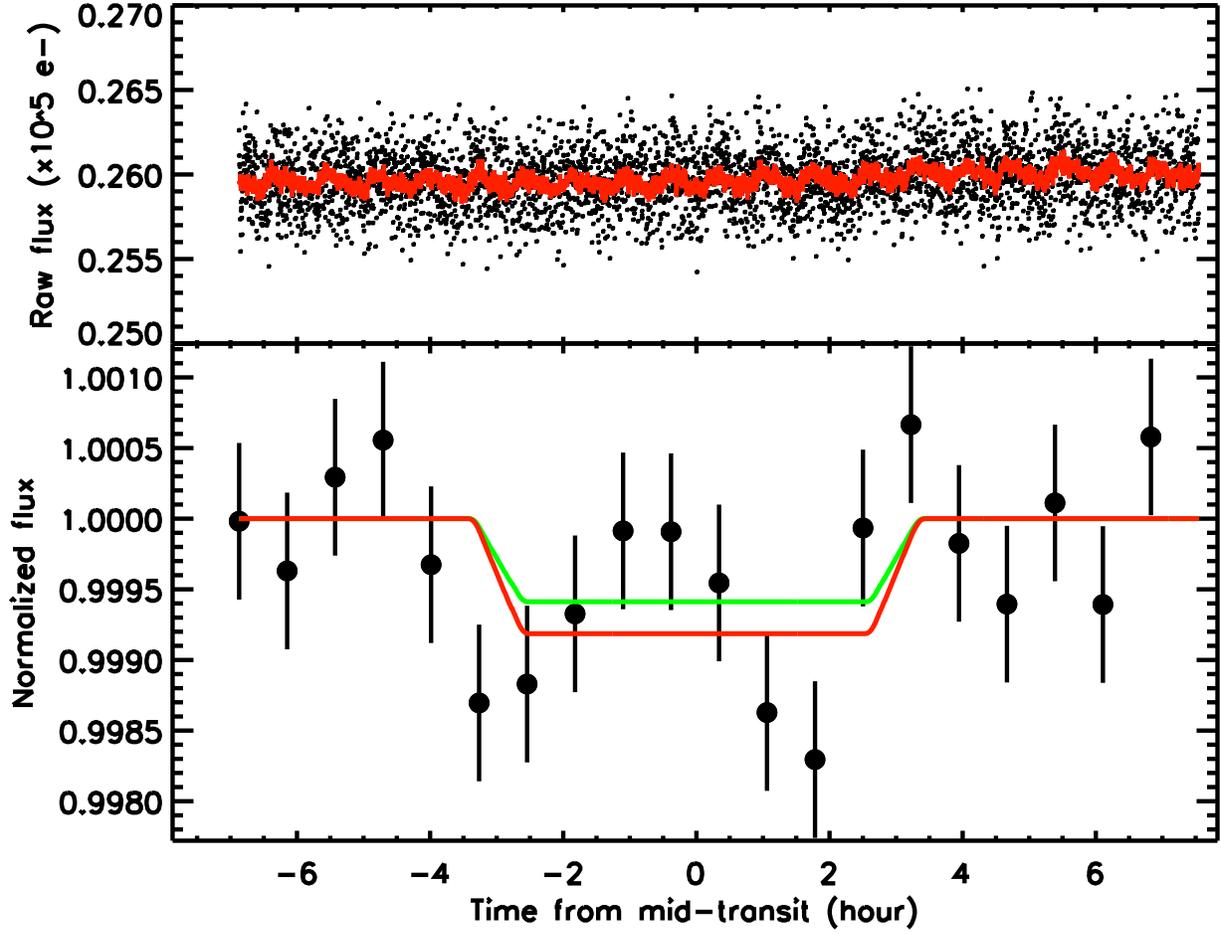}
 \caption{Example of a \spitzer\ transit light-curve observed in the IRAC band-pass at 4.5~\micron: KOI-701.03. Top panel: raw (unbinned) transit light-curve. The red solid line corresponds to the best fit model which includes the time and position instrumental decorrelations as well as the model for the planetary transit (see details in Sect.~\ref{sec:spfollowup}). Bottom panel: corrected, normalized and binned by 30 minutes transit light-curve with the transit best-fit plotted in red and the transit shape (with no limb-darkening) expected from the  \kepler\ observations overplotted as a green line. The two models agree at 1~$\sigma$ level.}
   \label{fig:spitzerlightcurves}
\end{center}
\end{figure*}

\begin{figure*}
\begin{center}
\setlength\fboxsep{17pt} 
\setlength\fboxrule{0pt} 
\fbox{\includegraphics[width=7.0in]{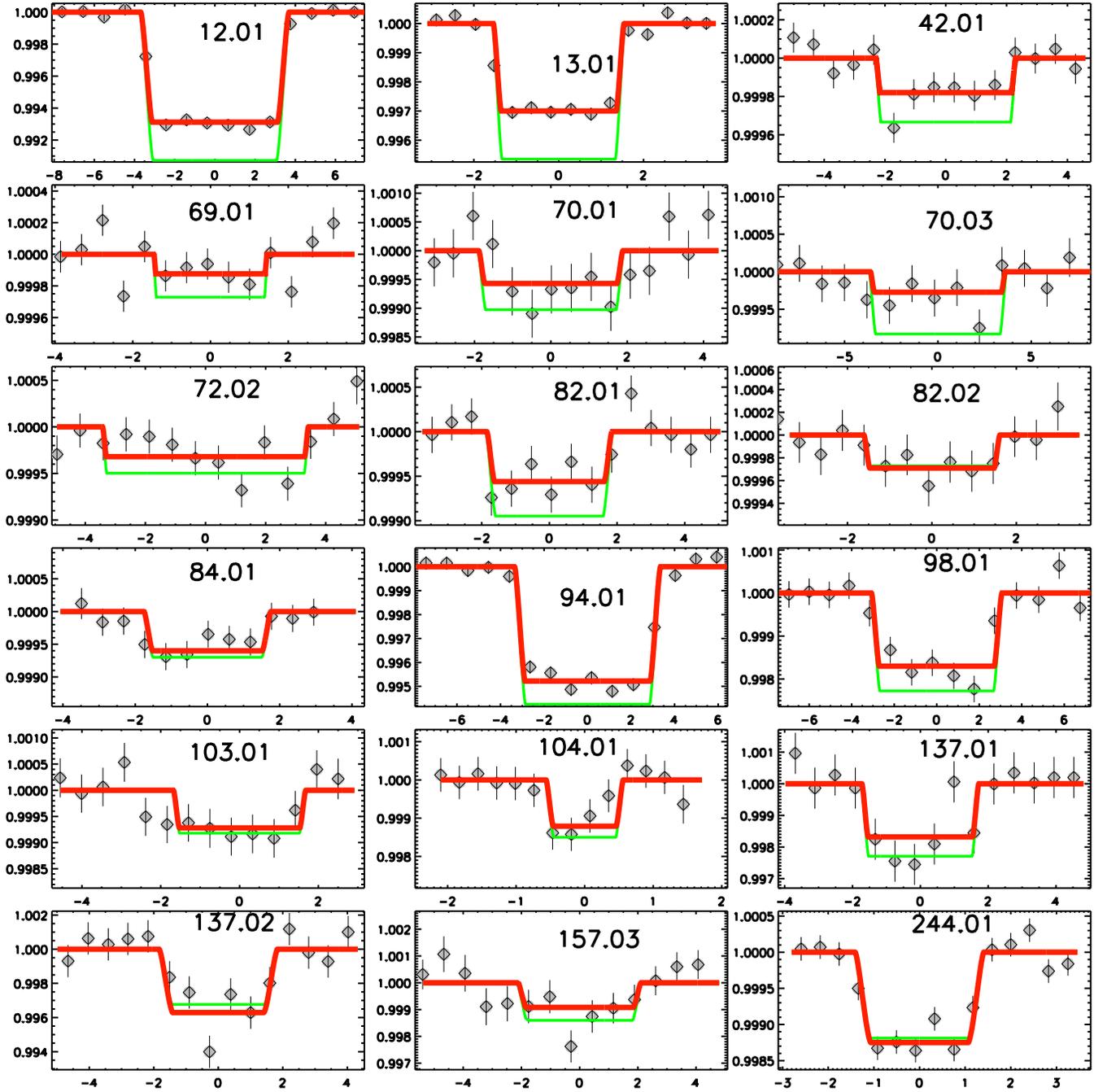}}
 \caption{Transit lightcurves from the \spitzer\ program 60028 (part-1/2). The lightcurves are obtained at 4.5~\micron\ with the IRAC instrument aboard \spitzer. The data are corrected, normalized, binned in time, and combined (when multiple observations are available). The grey points are the measurements with their 1~$\sigma$ error-bars. The red solid lines correspond to the best fit model of the \spitzer\ data (unbinned) as described in Sect.~\ref{sec:spfollowup}. The transit shapes expected from the \kepler\ observations are overplotted as green lines. The planetary transit models are computed neglecting the effect from stellar limb-darkening. The name of the KOIs appear in each individual plot.}
\label{fig:spitzerlightcurvesA}
\end{center}
\end{figure*}

\begin{figure*}
\begin{center}
\setlength\fboxsep{17pt} 
\setlength\fboxrule{0pt} 
\fbox{\includegraphics[width=7.0in]{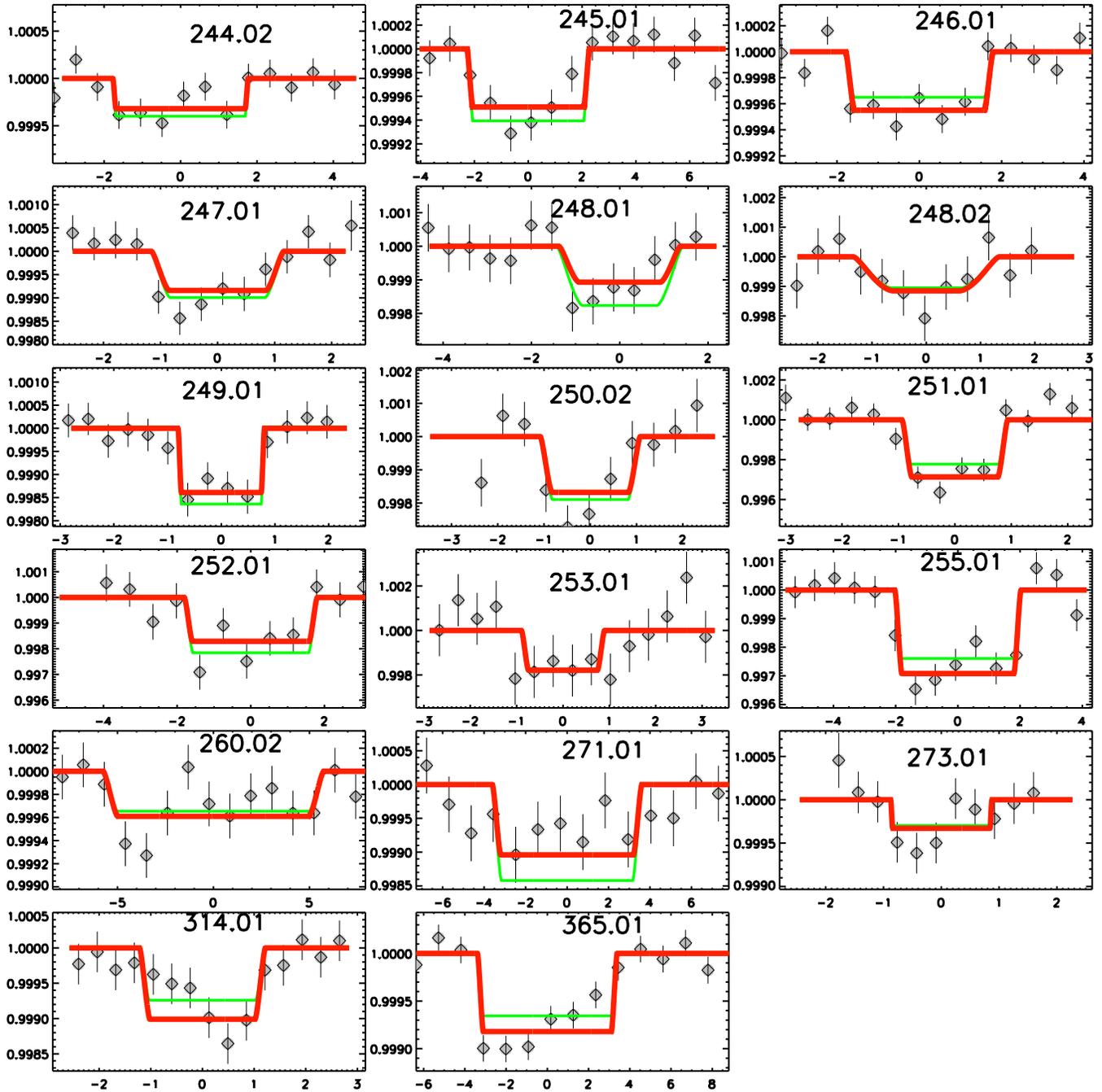}}
 \caption{Transit lightcurves from the \spitzer\ program 60028 (part-2/2). Same as Fig.\ref{fig:spitzerlightcurvesA}}
\label{fig:spitzerlightcurvesB}
\end{center}
\end{figure*}

\begin{figure*}
\begin{center}
\setlength\fboxsep{17pt} 
\setlength\fboxrule{0pt} 
\fbox{\includegraphics[width=7.0in]{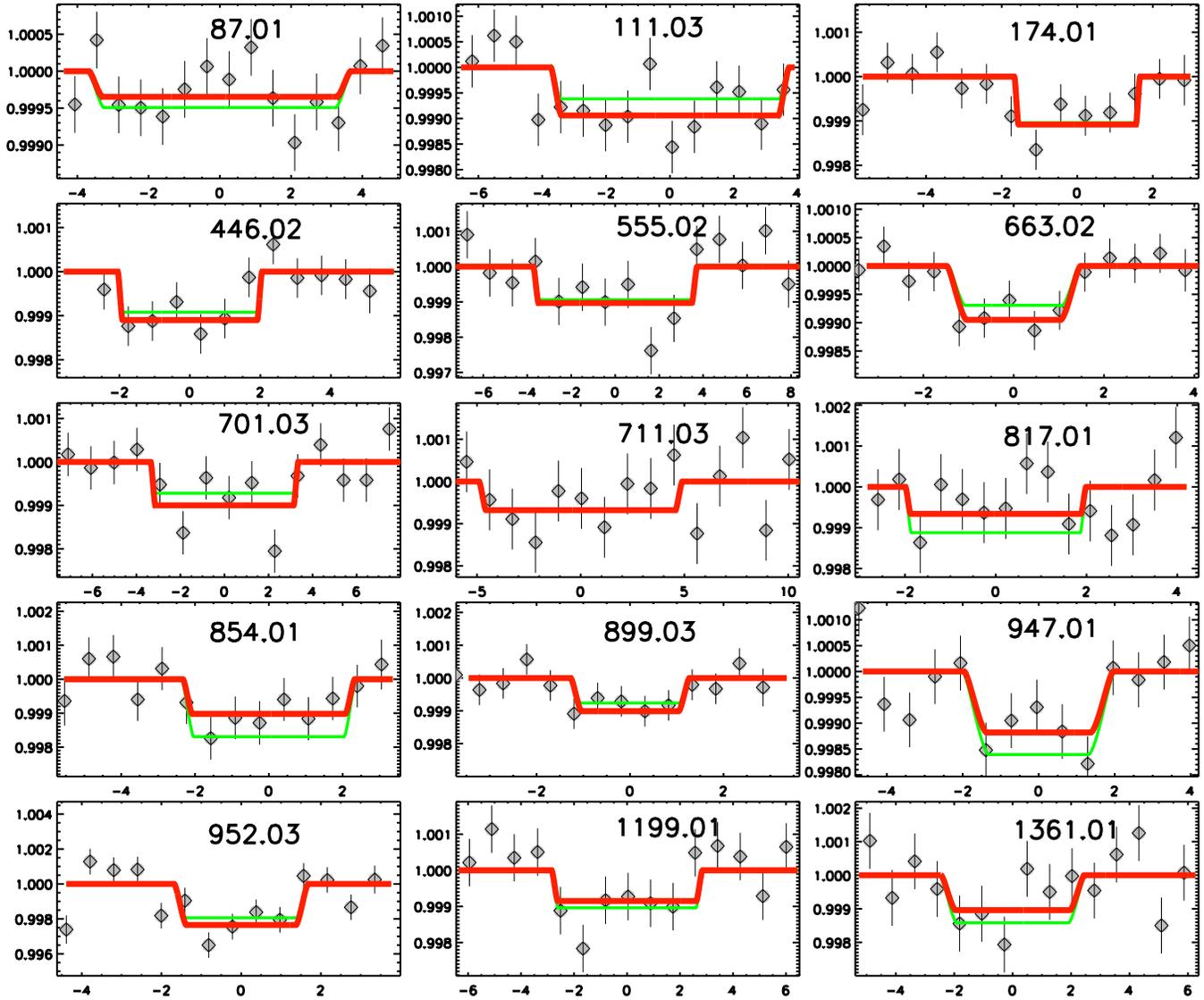}}
 \caption{Transit lightcurves from the \spitzer\ program 80117. Same as Fig.\ref{fig:spitzerlightcurvesA}}
\label{fig:spitzerlightcurvesC}
\end{center}
\end{figure*}

We perform a simultaneous Levenberg--Marquardt least-squares fit~\citep{Markwardt09} to the data to determine the transit depths and the instrumental parameters.
For each visit, we adopt the decorrelation function that significantly improves the $\chi^{2}$ minimization.
We rescale the errors on each photometric point to be set to the root-mean-square (rms) of the residuals from the initial best-fit of the data.
Hence, the reduced $\chi^{2}$ becomes one.
All the data-point measurement errors are therefore assumed to be identical for each lightcurve.
As an example, Figure~\ref{fig:spitzerlightcurves} shows the raw data and the corrected \spitzer\ transit lightcurve of KOI-701.03 (Kepler-62e). Figures~\ref{fig:spitzerlightcurvesA},~\ref{fig:spitzerlightcurvesB}, and ~\ref{fig:spitzerlightcurvesC} present the normalized, corrected, binned and combined lightcurves with their associated best fit models for all the observed KOIs that are presented in the current study.

We estimate parameter uncertainties using two different methods: Markov Chain Monte-Carlo (MCMC) and residual permutation methods.
Our MCMC implementation uses the Metropolis-Hasting algorithm with Gibbs sampling \citep{Tegmark04,Ford05}.
We assume uniform prior distributions for all jump parameters.
We adjust the width of the distribution from which we randomly draw the jump sizes in each parameter until 20--25\% of jumps are executed in each of the parameters.
We create five chains, each with 10$^{5}$ points, where each chain starts with a different set of starting parameters (each parameter is assigned a starting position that is +$3\sigma$ or $-3\sigma$ from the best-fit values).
We discard the first 10\% of jumps of each chain to remove the chain's transient dependence on the starting parameters.

In order to obtain an estimate of the correlated and systematic errors in our measurements, we use the residual permutation bootstrap, also called ``Prayer Bead'' method, described in~\citet{Desert11a}.
In this method, the residuals of the initial fit are shifted systematically and sequentially by one frame, and then added to the transit light curve model before fitting again.

For both methods, the posterior distributions are used to estimate the errors: we allow asymmetric error bars spanning $34\%$ of the nearest points above and below the values of the parameters associated with the minimum $\chi^{2}$ to derive the $1~\sigma$ uncertainties for each parameter.
We find that the two approaches provide consistent results.
Tables~\ref{tab:spitzer1} and~\ref{tab:spitzer2} present the transit depths and associated errors derived from the MCMC technique.

Finally, we check that KOIs for which we have multi-epoch measurements have transit depths that agree within the $3~\sigma$ level.
We combine the measured transit depths for these KOIs by computing the weighted means and errors.

\subsection{Determination of the \spitzer\ magnitudes}
\label{sec:spmag}

We use standard aperture photometry of each individual BCD image in order to compute the flux for all the KOIs in our sample.
We measure the averaged flux over the background annulus.
The main difference between the procedure used in the section and the one described in Section~\ref{sec:spdepth} is that we use a fixed aperture size with a radius of 3 pixels surrounded by an annulus of 12-to-20 pixels to estimate the flux and the background, respectively.
Furthermore, only the out-of-transit data are considered for determining the source flux densities.
We first convert the BCD images into mJy per pixel units from their original MJy per steradian units.
Then we estimate the centroid position of the main target star's PSF for each image, and compute aperture photometry centered on the source.
We apply an aperture correction of 1.113 at 4.5\micron\ (this value is taken from IRAC data handbook\footnote{{\tt http://irsa.ipac.caltech.edu/data/SPITZER/docs/irac/\\iracinstrumenthandbook/28/}}).
We correct the full lightcurve for the intrapixel sensitivity using the method described in Section~\ref{sec:spdepth}. 
Color and array-location-dependent photometric corrections are also applied to the photometry; the latter accounts for the variation in pixel solid angle (due to distortion) and the variation of the spectral response (due to the tilted filters and wide field-of-view) over the array~\citep{Hora04}.
Most of our data taken in full-array mode is such that the PSF is centered on the central pixel of the array (128;128), so no array correction is applied for this dataset.
The brightest stars of our sample are secured in subarray mode for which we apply an array correction of 0.68\%.
We convert the surface brightness (in mJy) into \spitzer\ magnitudes at 4.5\micron\ using a zero-magnitude flux density (zmag) of 179.7 Jy as computed by~\cite{Reach05}.
We finally compute the uncertainties on the flux densities using  the photon noise and the standard deviation of the measurements from all the individual frames.
We test the accuracy of our procedure using \spitzer\ IRAC photometric calibrator datasets taken from the public \spitzer\ archive: BD+60 1753 and HD180609.
We check that our magnitudes match those of~\cite{Reach05} at better than the $1~\sigma$ level.
Tables~\ref{tab:spitzer1} and~\ref{tab:spitzer2} present the flux densities and the corresponding magnitude at 4.5~\micron\ for each KOI.
While the uncertainties in these tables are the formal values, we have conservatively adjusted some of the errors to be no lower than 2\%. This lower limit is based on the findings of ~\citet{Reach05}.


\section{Using \spitzer\ observations to rule-out false positive scenarios}
\label{sec:spconstraint} 

\subsection{Analytical framework}

This section describes in detail how we use \spitzer\ observations to rule-out false positive scenarios.
The applied methodology makes use of the transit depths measured with \spitzer\ and with \kepler, as well as the measured colors \kepler --\spitzer.

The true transit depth $\delta_t$ obtained from an eclipsing system comprising a main object (2) and its companion (3) corresponds to:

\begin{equation}
\label{eq:tdepth}
\delta_{t} = \displaystyle \frac{\delta F_2} {F_2+F_3},
\end{equation}

\noindent where $F_2$ and $F_3$ are the emitted fluxes in the same bandpass. The parameter $\delta$ represents the surface ratio between the two objects and is expressed as:

\begin{equation}
\label{eq:depth}
\delta = \left(\frac{R_3}{R_2}\right)^{2}.
\end{equation}

The blended transit depth corresponds to the apparent transit depth $\delta_b$ of this eclipsing system diluted with a primary star (1). It is computed as follows:

\begin{equation}
\label{eq:bdepth}
\delta_{b} = \displaystyle \frac{\delta F_2} {F_1+F_2+F_3} = \delta_{t} \cdot d,
\end{equation}

\noindent where $F_1$ is the flux from the primary star, the targeted KOI, and $d$ is the dilution due to the presence this star in the line of sight of the eclipsing binary.

\noindent The ratio of the apparent transit depths measured in the \kepler\ (K) and the \spitzer\ (S) bandpasses corresponds to:

\begin{equation}
\label{eq:ratiodepth}
\frac{\delta_{b,{\rm S}}} {\delta_{b,{\rm K}}} = \displaystyle \frac{\delta_{t,{\rm S}}} {\delta_{t,{\rm K}}} \cdot \frac{d_{S}} {d_{K}},
\end{equation}

\noindent The ratio of the true transit depths is calculated from Eq.~\ref{eq:tdepth} and is expressed as:

\begin{equation}
\label{eq:ratiotruedepth}
\frac{\delta_{t,{\rm S}}} {\delta_{t,{\rm K}}} = \displaystyle \frac {F_{2,{\rm S}}/F_{2,{\rm K}} } { (F_{2,{\rm S}}+F_{3,{\rm S}} )/(F_{2,{\rm K}}+F_{3,{\rm K}})}.
\end{equation}

In order to simplify the problem, we assume that the contribution of the tertiary flux to the ratio of the dilution in the two bandpasses is negligible.
Omitting the contribution of the tertiary has a similar effect as reducing the distance modulus.
Therefore, this approximation does not significantly impact the final results.
Under this assumption, the ratio of the dilution in the \spitzer\ bandpass to that in the \kepler\ bandpass can then be written as:

\begin{equation}
\label{eq:ratiodilut}
\frac{d_{S}} {d_{K}} = \displaystyle \frac{10^{-0.4(\mathrm{M}_{1,{\rm K}} - \mathrm{M}_{2,{\rm K}} + \mu )}+1} {10^{-0.4(\mathrm{M}_{1,{\rm S}} - \mathrm{M}_{2,{\rm S}}  + \mu)}+1} ,
\end{equation}

\noindent where M$_{1}$ and M$_{2}$ are the absolute magnitudes of the primary and the secondary stars. The distance between the binary and the primary star is parametrized for convenience in terms of the difference in distance modulus, $\mu$ ($\mu=0$ for HT scenarios).  

Finally, the three component ``\kepler--\spitzer'' color $C_{\mathrm{KS}_{123}}$ is expressed as:
\begin{eqnarray}
\label{eq:color}
C_{\mathrm{KS}_{123}}&=&-2.5\times \nonumber \\
& & \log \frac{10^{-0.4(\mathrm{M}_{1,{\rm K}}+\mu
    )}+10^{-0.4(\mathrm{M}_{2,{\rm K}})}+10^{-0.4(\mathrm{M}_{3,{\rm K}})}}{10^{-0.4(\mathrm{M}_{1,{\rm S}}+\mu
    )}+10^{-0.4(\mathrm{M}_{2,{\rm S}})}+10^{-0.4(\mathrm{M}_{3,{\rm S}})}}. \nonumber \\
\end{eqnarray}

\subsection{Using the analytical framework to explore blend effects}

As a first step, we apply the above analytical framework to reveal the effect of blends in various scenarios that we expect to encounter. 
In particular, we explore blend types involving EB and HT systems that can mimic the KOI properties in both \kepler\ and \spitzer\ bandpasses.

We simulate blends using model isochrones from the Padova isochrone series~\citep{Girardi02}. 
This allows us to set the properties of the three stars involved, specifically their masses, from which their brightnesses can be predicted in any bandpass. 
The relevant bandpasses for this work are those that correspond to \kepler\ and \spitzer. 
These bands are available on the Padova models's CMD website\footnote{http://stev.oapd.inaf.it/cgi-bin/cmd}. 
For simplicity, we adopt here a representative isochrone of 3\,Gyr and solar metallicity.
The validity of this approximation, which has only a minor impact on our final results, is discussed in Sect.\ref{sec:complementstudies}.

\begin{figure*}[h!]
\begin{center}
\includegraphics{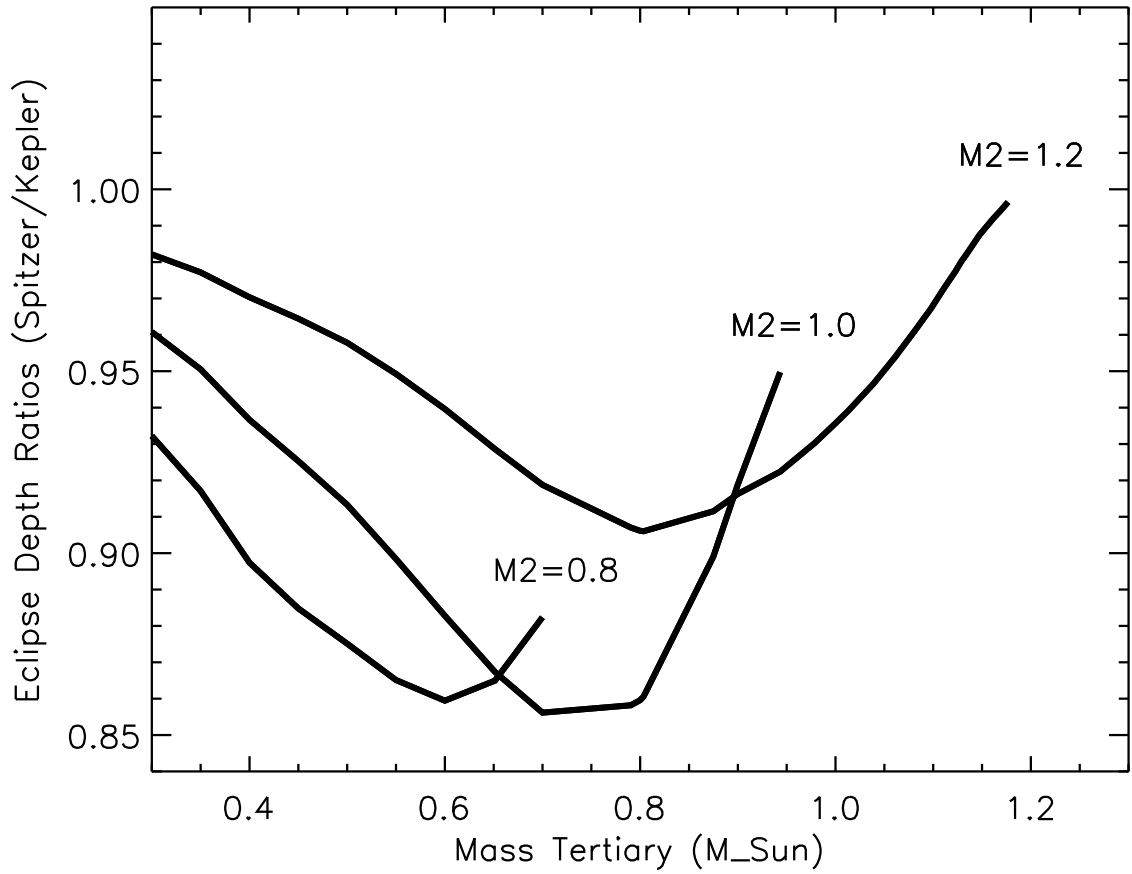} 
\caption{Ratios of the eclipse depth integrated in the \spitzer\ photometric bandpass over the eclipse depth integrated in the \kepler\ photometric bandpass as function of the mass of the tertiary star ($M_{3}$). These ratios are computed for an eclipsing binary stellar system composed of a secondary star of mass $M_{2}$ eclipsed by a tertiary stellar component (see Equation~\ref{eq:ratiotruedepth}). }
  \label{fig:ratiodepth}
\end{center}
\end{figure*}

\begin{figure*}[h!]
 \includegraphics{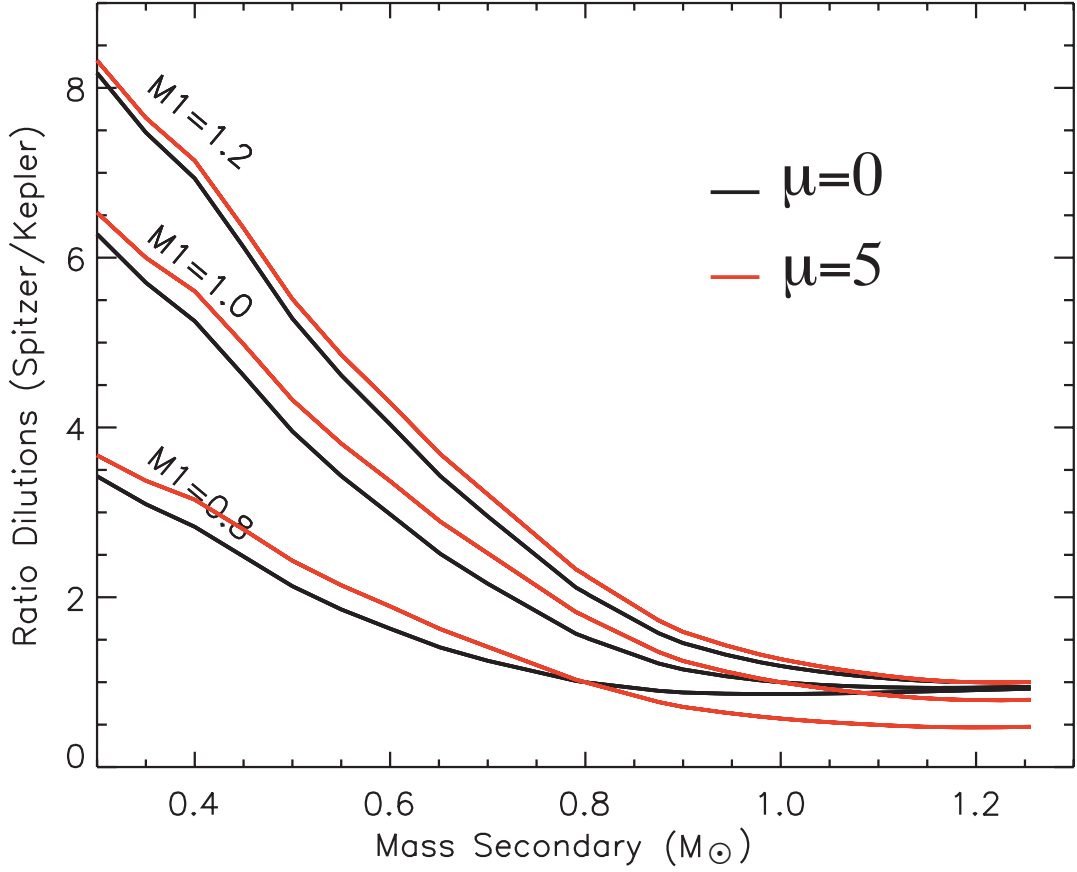}
 \caption{Ratios of the dilutions integrated in the \spitzer\ photometric bandpass over the dilution integrated in the \kepler\ photometric bandpass as a function of the mass secondary star ($M_{2}$) for an eclipsing binary stellar system. This system is composed of a secondary star eclipsed by a tertiary stellar component, that is blended with a primary star of mass $M_{1}$. Two scenarios of distance modulus $\mu$ are presented: $\mu=0$, implying that the binary system is equidistant to the primary star (HT: hierarchical triple), and $\mu=5$ for a background binary (EB: eclipsing binary) scenario. Three cases of $M_{1}$ are presented for reference (see Equation~\ref{eq:ratiodilut}).}
\label{fig:ratiodilut}
\end{figure*}

\begin{figure*}[h!]
\begin{center}
\includegraphics{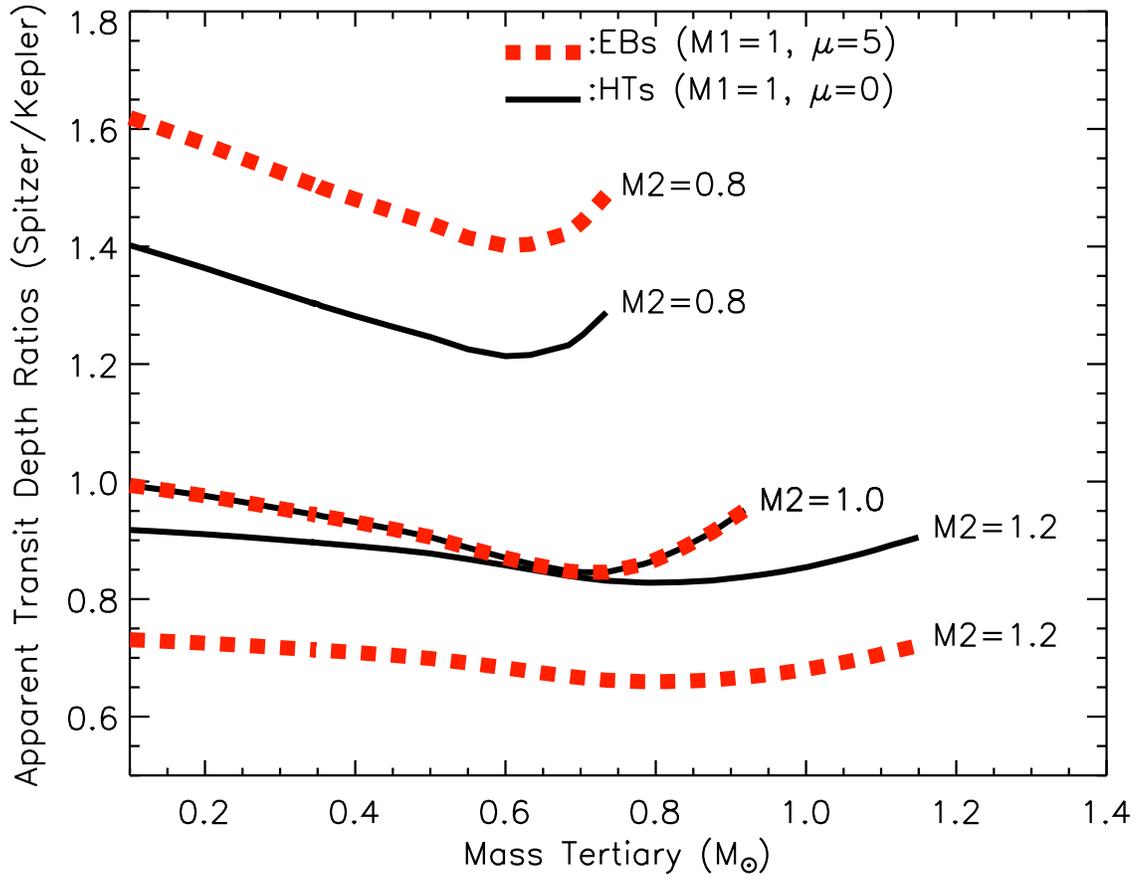}
 \caption{Same as Figure~\ref{fig:ratiodepth}, but the eclipsing binary system is now blended with a primary star. These ratios are computed for an eclipsing binary stellar system composed of a secondary star of mass $M_{2}$ eclipsed by a tertiary stellar component that is blended with a primary star of solar mass ($M_{1}$=1). See Equation~\ref{eq:ratiodepth} for more details.}
  \label{fig:ratiodepthd}
\end{center}
\end{figure*}

\begin{figure*}[h!]
\begin{center}
\includegraphics{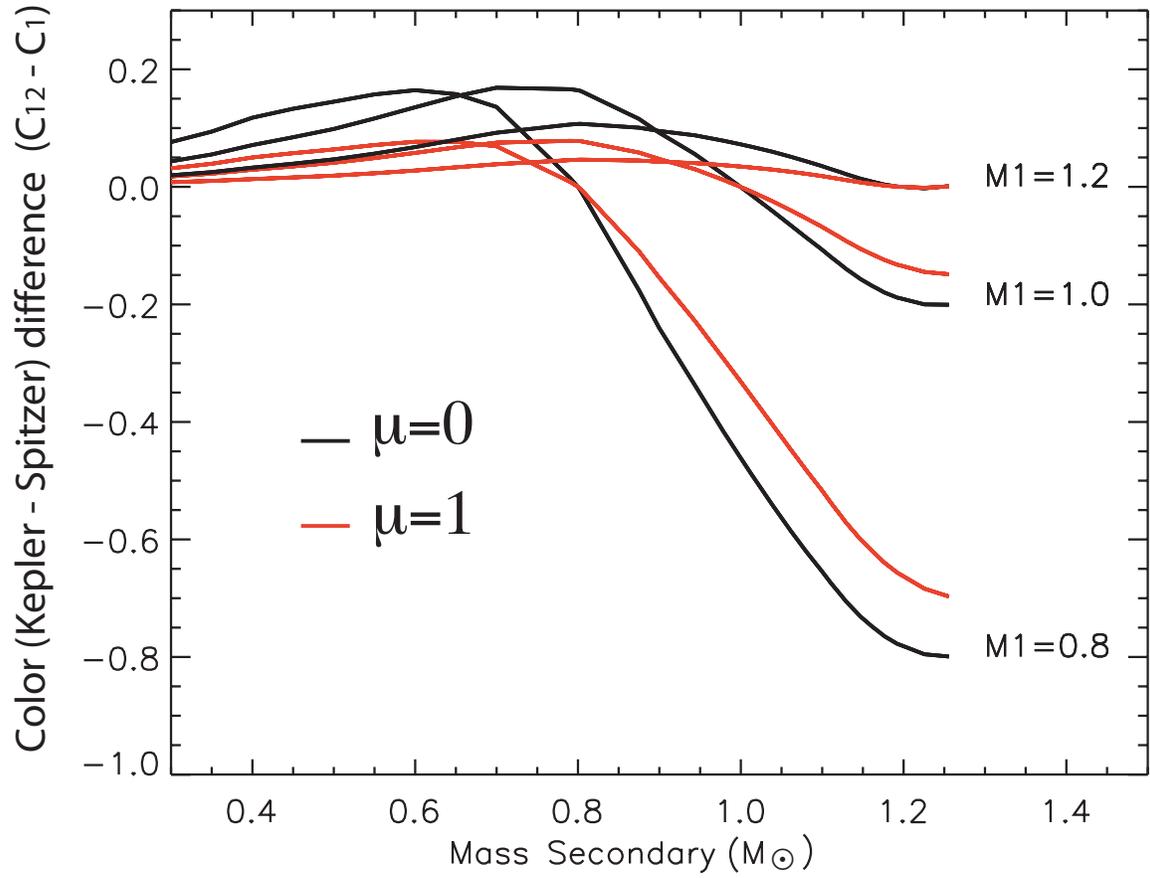}
 \caption{ Color (\kepler- \spitzer) difference between the combination of the two stars (primary and secondary) as a function of the mass $M_{2}$ of the secondary. The calculations are presented for three mass scenarios for the primary star. The black curves represent the color differences for a secondary star equidistant to the primary (HT scenario) and the red curves are for a background secondary at a distance modulus from the primary corresponding to 1~magnitude.}
  \label{fig:color}
\end{center}
\end{figure*}

We now present some of the characteristic features from our simulated blends, and we discuss how they change with the adopted stellar parameters. 
Figure~\ref{fig:ratiodepth} displays the ratio of the true eclipse depths in the \spitzer\ and \kepler\ bands for an undiluted eclipsing binary (eq.~\ref{eq:ratiotruedepth}) as a function of the mass of the tertiary star ($M_3$). 
We show this ratio for three different masses of the secondary star ($M_2$). 
The net dilution effect caused by the primary star as a function of secondary mass for three values of $M_1$ is presented in Figure~\ref{fig:ratiodilut}. 
Figure~\ref{fig:ratiodepthd} shows the same three cases as in Figure~\ref{fig:ratiodepth}, though we include this time the effect of the dilution in the signals produced by a primary star of one solar mass ($M_1 = 1\,M_{\sun}$). 
These apparent \spitzer/\kepler\ eclipse depth ratios (eq.~\ref{eq:ratiodilut}) are shown for two different blend scenarios: a HT configuration (in which the difference in distance modulus between the primary and the EB is zero), and a configuration with the EB in the background (a distance modulus of five). 
Finally, figure~\ref{fig:color} illustrates how the stellar properties affect the color difference of a blended system for colors computed between the \kepler\ and \spitzer\ bandpasses. 
This plot shows the difference between the color of the combined light of the primary and secondary ($C_{12}$) and the color of the primary alone ($C_1$), as a function of the secondary's mass. For this illustration, we assume that the tertiary's contribution is negligible, as will usually be the case.

\subsection{Applying the framework to the \spitzer\ and \kepler\ observations }

The methodology described above allows us to reject many false positive scenarios for the KOIs observed in this study.
The observational constraints consist of the measured transit depths and the apparent magnitudes in the \spitzer\ and \kepler\ bandpasses (see Tables 1 and 2).
The tests for potential blends use eqs.\ 5 and 6.
The free parameters considered are the secondary and tertiary masses, and the relative distance between the eclipsing pair and the primary star (i.e., the difference in distance modulus).
We explore these quantities over wide ranges of stellar masses and distance moduli, in a grid pattern to fully map the space of parameters for allowed blends.
Primary masses for the KOIs ($M_1$) are taken from the work of \citet{Batalha12} and held fixed.
We allow $M_2$ and $M_3$ to vary between 0.1\,$M_{\sun}$ and 1.4\,$M_{\sun}$, and the distance modulus difference $\mu$ is divided in linear steps between $-5$ mag and 15 mag. 
For each star, the intrinsic brightness in the \spitzer\ and \kepler\ bands is read off directly from the adopted isochrone. 
At each trial distance modulus we compute the ratio between the transit depth from \spitzer\ ($+ \, \, 3\sigma$) and that from \kepler.
This ratio sets an upper limit to the mass of the secondary that could mimic the transit depths measured in both banpasses.
We do not set a lower limit between the transit depths from \spitzer\ ($- \, \, 3\sigma$) and that from \kepler, because these shallower depths would involve scenarios with massive stars (these are typically more massive than the Sun; See Figure~\ref{fig:ratiodepthd}).
In practice, the number of stars that we eliminate from the shallower limits represent only a small fraction of the number of low mass stars removed from the deeper limits.
In a typical \kepler\ aperture, the Besan\c{c}on population synthesis model shows that only 10\% of the stars have mass greater than 80\% of the Sun.
Consequently, we only consider the deeper transit depth from \spitzer\ ($+ \, \, 3\sigma$) in our calculation of the FPPs.

Similarly, we use the \kepler\ {\it minus} \spitzer\ color to further constrain the blend properties.
The color is derived from our measured \spitzer\ magnitude (m$_{S}\pm3\sigma$ ; Table~\ref{tab:spitzer1} and~\ref{tab:spitzer2}) and from the \kepler\ magnitude as reported by \citet{Batalha12}.
Assuming that the measured fluxes result from the contribution of three stars, the color $C_{123}$ (eq.~\ref{eq:color}) is then compared with the color of the primary star alone, $C_{1}$, computed from the adopted isochrone. 
Only a subset of the secondary and tertiary stars can reproduce this color difference, which provides another constraint on the secondaries' masses.


\section{Additional Observational Constraints}
\label{sec:observations}

This section presents additional observational constraints used for several candidates of our sample. 
These complementary observations allow us to exclude more false positive scenarios, which remain after applying the constraints from the \spitzer\ observations.

\subsection{Stellar Reconnaissance}
\label{sec:recon}

In general, follow-up observations of \kepler\ planet candidates involve reconnaissance spectroscopy.
This is necessary to characterize the primary star and to look for evidence of astrophysical false positives~\citep{Batalha10a}.
Such false positives include single- and double-lined binaries, some HTs and EBs, which would show velocity variations at large amplitudes.
We also use their spectra to estimate the effective temperature, surface gravity, metallicity, and rotational and radial velocities of the host star.
For the current study, we assume that the primary star is the brightest star and that it is known and characterized.
In theory, if we are to consider all possible scenarios, a secondary star could be brighter than the primary target (see Figure~\ref{fig:ratiodepthd}).
In practice, our assumption that the primary star is characterized means that the secondary stars considered in this study can only be fainter than the primary stars.
We treat the stars in our sample in a uniform manner as we rely on the stellar radii and masses provided in~\cite{Batalha12}, which are from the KIC~\citep{Brown11}.
This is a critical step to fix the primaries' stellar characteristics in our FPP calculations.

\subsection{Imaging}

\kepler's photometric aperture is typically a few pixels across with a scale of 3\farcs 98 per pixel. 
Therefore, high-resolution imaging is often performed in order to identify neighboring stars that may be blended EBs contaminating the primary target photometry.
Only 23 amongst the 51 stars of our study have high spatial-resolution adaptive optics images.
These images were taken in the near-infrared (J and K bands) with ARIES~\citep{McCarthy98} on the MMT and PHARO~\citep{Hayward01} on the Palomar Hale 200-inch.
The observations and their sensitivity curves are presented in~\cite{Adams12}.
The AO images allow us to detect companion stars as close as 0.1\arcsec\ from the target's primary star.
These images also rule-out stellar companions within a 6\arcsec\ separation from the primary, with a magnitude difference up to 9.
There are several KOIs for which we detect additional stars within the \kepler\ apertures of the primaries.
We note that since we completed the analysis of the data for this work, several projects have been conducted to search for close-by stellar companions that could be the sources of false positives using high resolution imaging~\citep{Lillobox12,Adams13,Lillobox14}. 
Furthermore, a robotic AO survey of nearly 715 KOIs was conducted by~\cite{Law13}, to search for stellar companion in a systematic manner.
So far these searches found 7 KOIs within our samples that have detected fainter close-by stellar companions within 5\arcsec; these are KOI-12, 13, 94, 98, 111, 174, and 555.
We also note that the environment of KOI-854 has been scrutinized using HST/WFC3, but no companions were reported~\citep{Gilliland14}.

\subsection{Centroid analysis from \kepler\ }
\label{sec:centroids}

The very high astrometric precision of \kepler\ allows us to monitor the motion of the target's photocenter.
This provides an effective way of identifying false positives that are caused by EBs falling within the aperture.
We directly measure the source location {\it via} difference images to search for impostors based on scrutinizing pixels in the KOIs' aperture.
Difference image analysis is conducted based on the difference between average in-transit pixel images and average out-of-transit images.
A fit of the \kepler\ pixel response function (PRF;~\citealt{Bryson10}) of both the difference and out-of-transit images directly provides the transit signal's location relative to the host star.
The difference images are measured separately in each quarter.
We estimate that the transit source location is the robust uncertainty-weighted average of the quarterly results.
\cite{Jenkins10a} and \cite{Bryson13} describe this technique. 

Subsequent to our study, \cite{Bryson13} presented the centroid analyzes for the complete list of KOIs.
However, at the time of this study, the centroid analyzes have been performed only for a subset of the KOIs targeted in this work (see Table~\ref{tab:fpp}).
This analysis shows no significant offsets during transits in any quarter, the computed offsets are well within the radius of confusion (at the $3\sigma$ level).
This shows that the observed centroid locations are consistent with the transit occurring at the KOI locations.
Stars located at distances beyond the confusion radius from the targeted KOI are ignored since the centroid analysis would be confused by such stars and this would not yield accurate measurements. However, we note that bright stars can have PRF wings than can extend to over 100\arcsec\ and could therefore contaminate the main target~\citep{Coughlin14}.

\subsection{Faint limit condition}
\label{sec:faintlimit}

The faint limit condition corresponds to the faintest blended EB system than can reproduce the transit signal.
The blended system must comprise more than a fraction $\delta_{b}$ (as defined in Equation~\ref{eq:bdepth}) of the total flux within the \kepler\ aperture.
This condition may be expressed as the following:
 \begin{equation}
 \label{eq:blender}
m_{2\mathrm{lim},{\rm K}} -  m_{1,{\rm K}} = \Delta m_K = -2.5 \log_{10}(\delta_{b}),
\end{equation}
\noindent where $m_{2\mathrm{lim},{\rm K}}$ is the apparent \kepler\ magnitude of the blended binary system and $m_{1,{\rm K}}$ is the magnitude of the \kepler\ targeted star.  
This limit is such that no binary system fainter than $m_{2\mathrm{lim},{\rm K}}$ can possibly mimic a transit signal with a depth of $\delta$ given the primary star of \kepler\ magnitude $m_{1,{\rm K}}$. 


\section{False-Positive Probability}
\label{sec:FPP}

In this section, we compute the false positive probability for each KOI in our selected sample.
For any candidate, the rate of false positives relative to the rate of any event can be written as:
\begin{equation}
\label{eq:fpp}
FPP=\frac{F_{FP}}{F_{FP}+F_{P}},
\end{equation}
where $F_{FP}$ is the estimated frequency of false positive scenarios (which depends on the local stellar density), and $F_P$ is the expected frequency of transiting planets for a given KOI.
Stars that are unable to reproduce the observables are removed using complementary observational constraints (e.g., centroid analysis, AO imaging, etc...).
We then compute the planet and false positive frequencies as described in the following sections and we finally derive the FPP for each object.
We present the FPP for each KOI observed with \spitzer\ in Table~\ref{tab:fpp}.
Further details about the steps undertaken are given below.

\textit{\subsection{Planet Frequency}}

\begin{figure*}[h!]
\begin{center}
\includegraphics{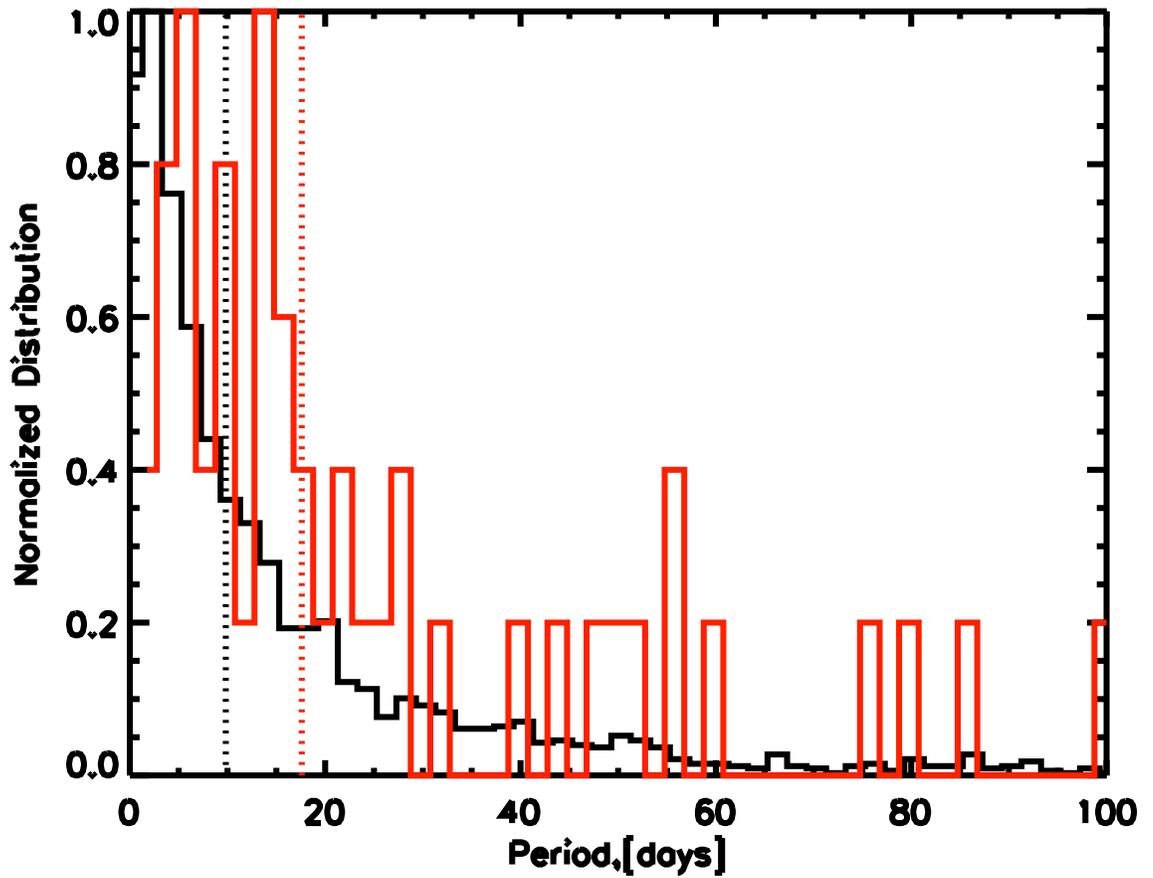}
 \caption{Normalized period distribution of the KOIs (in black) and the \spitzer\ sample (in red). The KOI distribution is derived from~\cite{Batalha12}. The vertical dashed lines correspond to the median value of each distribution. The two distributions are broadly consistent, except at the very shortest orbital periods.}
  \label{fig:period_distribution}
\end{center}
\end{figure*}

\begin{figure*}[h!]
\begin{center}
\includegraphics{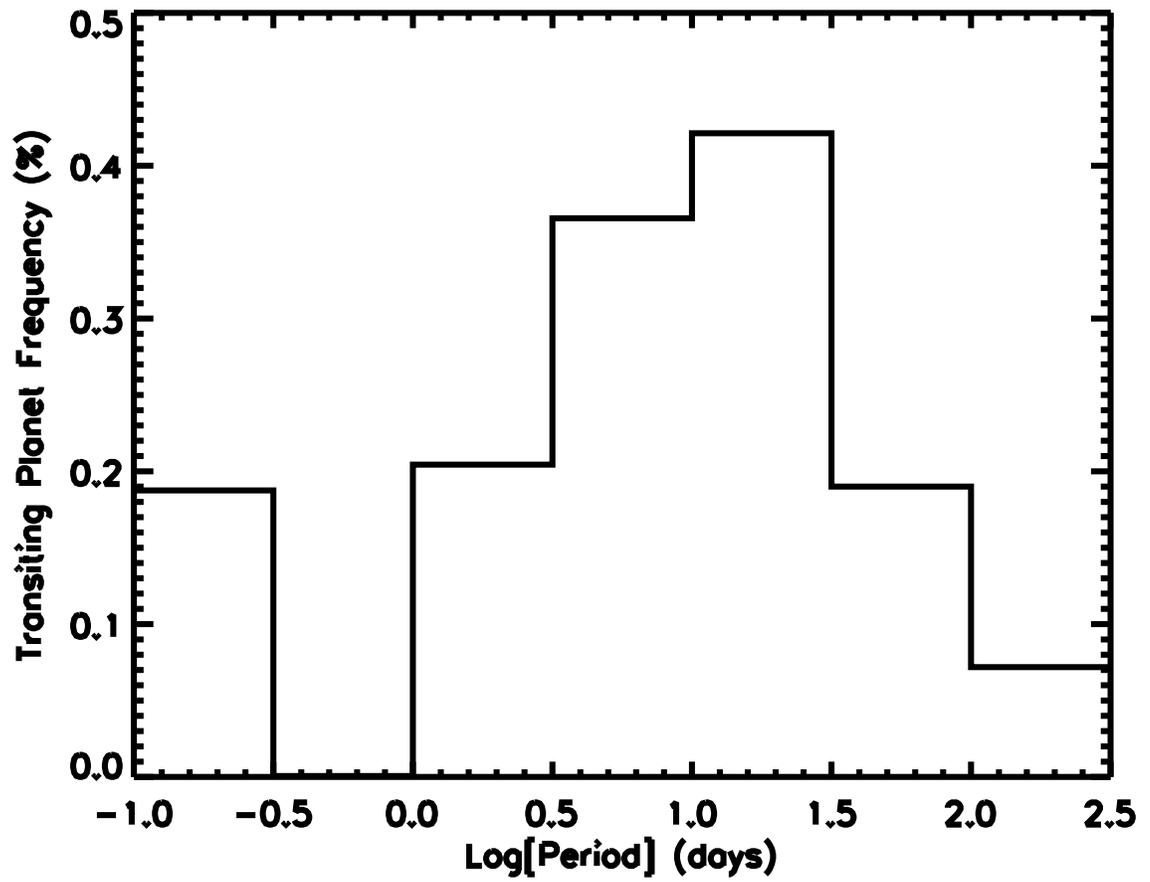}
 \caption{Transiting planet frequency distribution as function of period bins (in logarithmic scale). The distribution is derived from the list of KOIs published in~\cite{Batalha12}, assuming that all the detected signals are of planetary origin.}
  \label{fig:pf}
\end{center}
\end{figure*}

To estimate the likelihood of a planet, we rely on the list of candidates from Kepler.
We assume here that all the KOIs are true planets and that the period distribution of our \spitzer\ sample follows, in first order, the period distribution of the KOIs. 
The latest assumption is verified in Figure~\ref{fig:period_distribution} for the KOI distribution derived from~\cite{Batalha12}.
The KOIs are separated per period range using equal logarithmically spaced intervals.
We count the number of candidates that each period bin contains and divide by the total number of stars followed by \kepler\ (156,453) to derive the planet frequency for a given bin.
Figure~\ref{fig:pf} shows the distribution of planet frequencies that we use in the calculation of the FPP for the KOIs we observed with \spitzer.

\subsection{False-Positive Frequency}

In order to derive the $F_{FP}$ of a KOI, we must assess the likelihood of the various types of false positives, and also estimate the local stellar density.
We estimate the stellar density using a stellar population synthesis model of the Galaxy, the Besan\c{c}on model~\citep{Robin03}.
We use this model to derive the frequency of stars present in the photometric apertures around each KOI in our sample.
We adopt the stellar densities predicted by this model in the R band, which is a band sufficiently close to the \kepler\ bandpass for our purposes.
Instead of estimating the stellar population in a cone around the line of sight of each KOI, we create a grid of 24 cells evenly spaced over \kepler's field of view (about one cell per \kepler\ CCD module).
An aperture of 1~square degree centered on each cell of the grid is chosen for the initial estimate of stellar population.
We then perform stellar density calculations in half-magnitude bins of apparent brightness, from a R-magnitude of 5 down to magnitude 24. 
We also account for interstellar extinction with a coefficient of 0.5~mag~kpc$^{-1}$ in $V$ band.
The number of stars that fall in the aperture of each grid varies between 30,000 and 1,400,000, depending on the Galactic latitude and longitude of the KOIs.
We derive the stellar population using the closest cell relative to the target.
In this way, we evaluate the expected number density of neighboring stars, and their mass distribution per square degree and per magnitude bin.
We then estimate the number of EBs amongst these neighbor stars that could potentially reproduce the signal detected in \kepler.
It corresponds to the occurrence rate of binaries multiply by the probability that they undergo eclipses.
This rate has been measured in the \kepler\ field~\citep{Prsa11,Slawson11}.
In particular, 1261 detached binary systems have been found from the first four months of observations with periods less than 125~days~\citep{Slawson11}.
These are the typical false positives of interest for our study.
Therefore, the frequency of EBs that we consider is $f_{\rm EB} = 1261/156,\!453 = 0.80\%$.
This frequency is computed from short period detached binaries and it depends on their period; it decreases below 0.80\% for longer periods.
Consequently, the uniform frequency of EBs that we use in this work is a conservative value.
This allows us to compute upper limits for the FPPs.
We note that this frequency includes eclipsing pairs in triple or higher multiplicity stellar systems.
Finally, the total blend frequency $F_{FP}$ that is inserted in Equation~\ref{eq:fpp} is found by combining the probabilities associated with all background or foreground binaries.

\subsection{Combining observational constraints to remove blended stars within the \kepler\ apertures}

For each KOI in our sample, we use the density of stars in each magnitude bin determined using the population synthesis model. 
We now calculate the fraction of these stars that would remain false positive candidates after taking into account the available observational constraints.
We describe below the method used in this task. 

The first step uses the faint limit condition described in Section~\ref{sec:faintlimit}.
This removes all the stars that are too faint to reproduce the transit depths observed in the \kepler\ photometry.
The remaining stars pass through the second step that combines high-resolution images of the target's neighborhood (AO observations) and centroid motion analyzes when available.
This provides the spatial extent considered for blend frequency calculations.
These constraints significantly decrease the size of the apertures that could hide a background or foreground binary: this is sometimes 5'' and often 2''.
Even though stars located at distances beyond 5'' can reproduce \kepler\ signals~\citep{Coughlin14}, we assume for simplicity that all the stars that are beyond the limits found by the centroid method cannot mimic transit signals.

These stars are therefore excluded from the blend frequency calculations.
This step always removes the largest number of possible blend scenarios.
It is a very powerful tool for identifying background EBs blended with the target~\citep{Batalha10a}.
In a final third step, we use the constraints from the \spitzer\ observations, the measured transit depths and magnitudes, as described in Section~\ref{sec:spconstraint}.
For each secondary star that survives the first two steps, we consider tertiary stars (EBs), ranging from 0.1 to 1.4 solar masses, and test the ratios of transit depths.
We compute the diluted ratio of the true transit depths for each scenario following Equation~\ref{eq:ratiodepth}. 
We compare this ratio to its observed value for each mass of the chosen tertiary star. 
If the calculated ratio is not consistent with the observed one, the tertiary star scenario is then rejected. 
We also apply the constraint from the color ``\kepler--\spitzer'' $C_{\mathrm{KS}_{123}}$ as given by Equation~\ref{eq:color}.
We reject the tertiary star scenarios for which the calculated colors are not in agreement with the measured ones.
Overall, this third step allows us to remove most of the red dwarfs that could potentially remain as false positives.
We finally find the total blend frequency by combining the probabilities associated with all background or foreground star.


\section{Discussion}
\label{sec:discussion}

\subsection{Comparing the \kepler\ and \spitzer\ transit depths}

\begin{figure*}[h!]
\begin{center}
\includegraphics[width=6in]{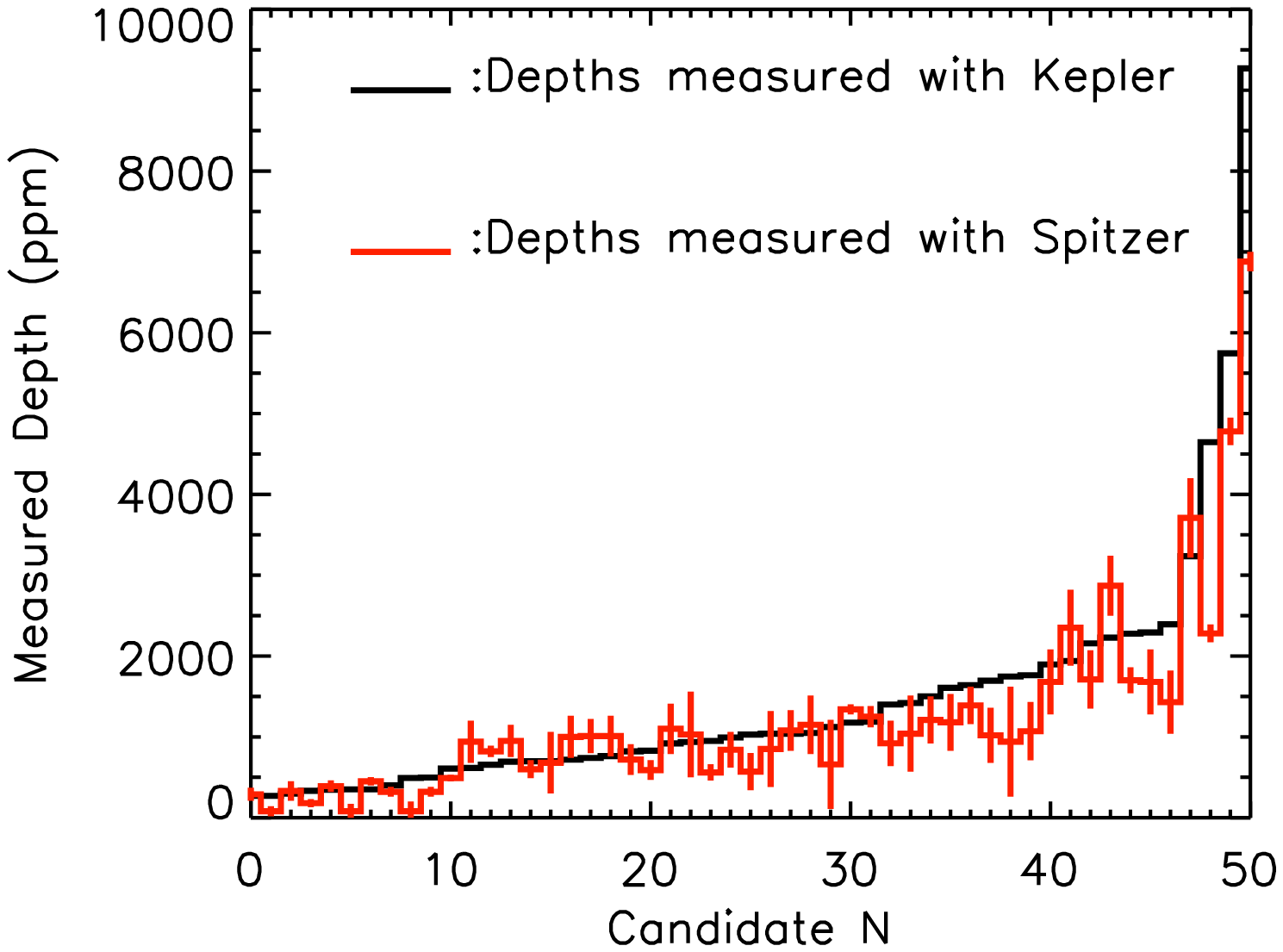}
\includegraphics[width=6in]{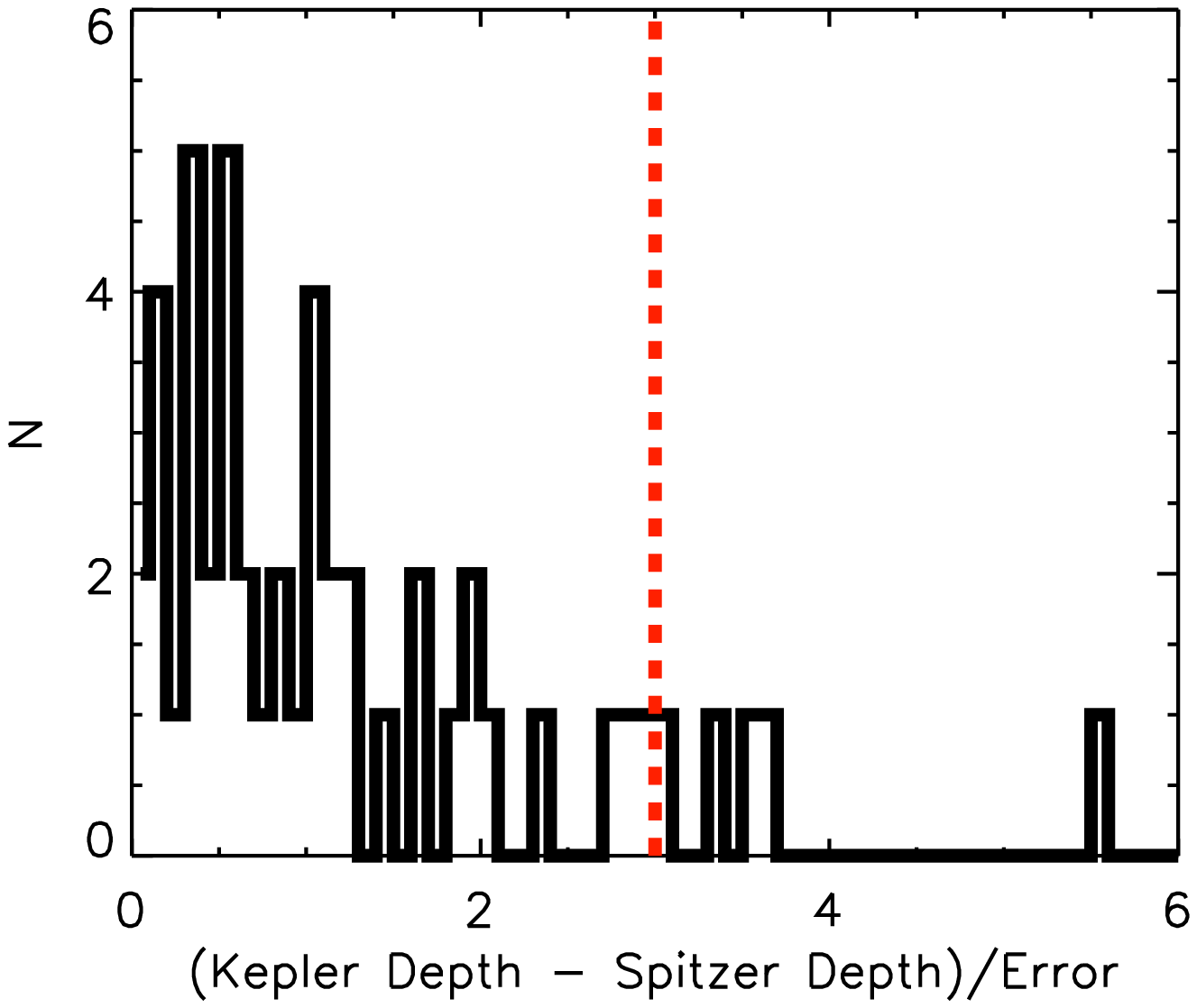}
\caption{{\bf Top:}  Transit depths measured from the \spitzer\ 4.5~\micron\ lightcurves (red histogram) compared to their depths measured in the \kepler\ bandpass (black histogram) for 50 KOIs targeted in this program. The error bars on the \spitzer\ measurements correspond to $1~\sigma$ uncertainties. The targets are ordered by increasing transit depths measured from \kepler\ towards the right. {\bf Bottom:} Distribution of the significance of the apparent transit depth differences measured between \kepler\ and \spitzer. The vertical dotted red line highlights the $3~\sigma$ uncertainties; it encompasses $85\%$ of the candidates. Two KOIs with differences greater than $6~\sigma$  do not appear on this figure (KOI-12.01 and 13.01). The difference in the apparent transit depths is not corrected for dilution caused by the presence of a close-by companion.}
\label{fig:depth}
\end{center}
\end{figure*}

Figure~\ref{fig:depth} (top panel) shows the measured transit depths in the \spitzer\ bandpass compared with those measured with \kepler.
We find that 50\% of the sample have measured depths that agree within $1\sigma$, and that 85\% agree within $3\sigma$ (Figure~\ref{fig:depth}, bottom).
The distribution is therefore somewhat broader (by $\sim$20\%) than expected for a Gaussian with a standard deviation of unity, indicating that the \spitzer\ and \kepler\ transit depths for our KOIs are not all statistically consistent within their uncertainties. This may be caused by: (i) the presence of false positives in our sample; (ii) dilution from unresolved companion stars resulting in wavelength-dependent transit depths; or (iii) underestimated uncertainties in our measurement of the Spitzer transit depths. Below we present evidence that some of the KOIs do indeed suffer from dilution effects. Likewise, biases in determining reliable uncertainties for our \spitzer\ measurements cannot be entirely ruled out, as they depend on our ability to correct the data for the main source of systematic errors, which is the intra-pixel sensitivity.

There are three candidates for which the transit depths measured from the \kepler\ photometry are significantly deeper ($>4~\sigma$) than the depths measured from \spitzer: KOI-12.01, KOI-13.01, KOI-94.01.
These objects are three Jupiter-size planet candidates out of the four from this family that we have in our sample.
Since we observed KOI-13.01 and KOI-94.01 with \spitzer, these candidates have been confirmed as {\it bona fide} planets by \citet{Barnes11}, and \citet{Weiss13} respectively.
These three KOIs have known close-by companions which are located at closer than 1'', well within the \kepler\ apertures and within the \spitzer\ PSFs.
The host star KOI-13 is known to be part of a stellar binary system, both components being rapidly rotating A stars \citep{Szabo11, Barnes11}. 
The companion is about 300~K cooler than the primary host star and is 0.3 magnitude fainter in the Kepler bandpass.
Similarly, the CFOP shows that KOI-12 is also a massive fast rotator star. Direct images of the close environment of KOI-12 reveal the presence of two fainter stars within 1'' of the primary host.
A low mass-star companion at 0\farcs 6 from KOI-94 has been detected and this explains the significant difference of measured transit depths~\citep{Takahashi13}.
The candidates with the largest discrepancies are KOI-82.01 and KOI-98.01 at the $3.7$ and $3.6 \sigma$ level, respectively. For KOI-98.01, the host star is known to have a stellar companion at  0\farcs 3 \citep{Buchhave11,Law13}. KOI-82 has no close by companion detected by AO~\citep{Marcy14}; we therefore attribute the discrepancy between \kepler\ and \spitzer\ to statistical fluctuations.
In general, the dilution produced by the presence of stars within the aperture of the KOI result in chromatic differences between the transit depths.
In the present cases, the flux contamination from the companion stars to the primary host targets, KOI-12, 94 and 13, vary with wavelength. The dilution is greater in the infrared compared to the visible as expected for cooler (redder) contaminants.
Because the contaminant stars contribute proportionally to more flux in the infrared compared to the visible, the dilution produced by their presence is larger at longer wavelengths.
This results in shallower measured transit depths in the \spitzer\ bandpass compared to the \kepler\ bandpass.

KOI-258.01 is the only observed candidate for which the lightcurve is not properly fitted by a transit planet model.
The \spitzer\ and the \kepler\ transit lightcurves exhibit a clear V-shape.
Such objects do not usually appear in the KOI list as they are flagged as false positives.
KOI-258.01 was ranked as a KOI early on in the mission before being removed.
The \kepler\ project has since changed this target to an inactive mode as the primary star has two companions within 1.5" that are 2 to 3 magnitudes fainter in the K band.
However, by the time of this decision the order to execute the \spitzer\ observation had already been given. For consistency we therefore choose to remove this object from the list of KOIs with computed FPPs.

\subsection{On the False Positive Rate}
\label{sec:onfpp}
As described in Section~\ref{sec:FPP}, we estimate the FPP for each individual candidate on this list and Table~\ref{tab:fpp} presents our results.
We find that half of the targets in our sample have a FPP which is lower than 1\% (see Fig.~\ref{fig:fpp}).
Using the distribution of FPP for the complete sample, we measure a median value of 1.3\%. 
We calculate a robust estimate of the dispersion of the FPP distribution using the median absolute deviation as the initial guess, and then weighting points using Tukey's Biweight~\citep{Hoaglin1983}.
The dispersion of our distribution measured by this method is 2.5\%.
This leads to an upper limit for the FPR of 3.8\% at the $1 \sigma$ level, and of 8.8\% at the $3 \sigma$ level.

At present, the sample of this study represents only 2\% of the total candidates published so far~\citep{Batalha12}.
Therefore, extrapolating a \kepler\ FPR from such a small sample should be done with caution.
Furthermore, our sample is not a uniform sample of KOIs as it can be divided into two categories. 
The first category comprises the sub-sample of two thirds of the targets for which we have complementary observational constraints from direct imaging (AO) or \kepler's centroid measurements.
For this first sample, we measure a median FPR of 0.7\% with a robust estimate of the dispersion of 0.8\%.
The second category comprises the rest of the targets for which we have no other constraints other than the \kepler\ depths and the \spitzer\ photometry.
This is because we do not have the centroid analysis completed yet at the time of the work.
For this second sample, we measure a FPR lower than 24\% at $1 \sigma$ level, and we find that these FPPs vary between 5-to-43 \%.
The KOIs of our sample with the highest FPP values (e.g. KOI-137.01, 137.02 and 952.02) are indeed confirmed planets, which implies that their FPP is much lower than the values we can compute from the current study.
All of this leads towards a rather low value for the overall FPR of the \kepler\ sample.
Finally, we note that 11 of the observed KOIs have FPPs $<$ 0.3\%.
This implies that these detected signals are at least 99.7\% consistent with planetary objects; these KOIs can be considered as ``validated'' at the $3 \sigma$ level of confidence.

\begin{figure*}[h!]
\begin{center}
\includegraphics[width=6in]{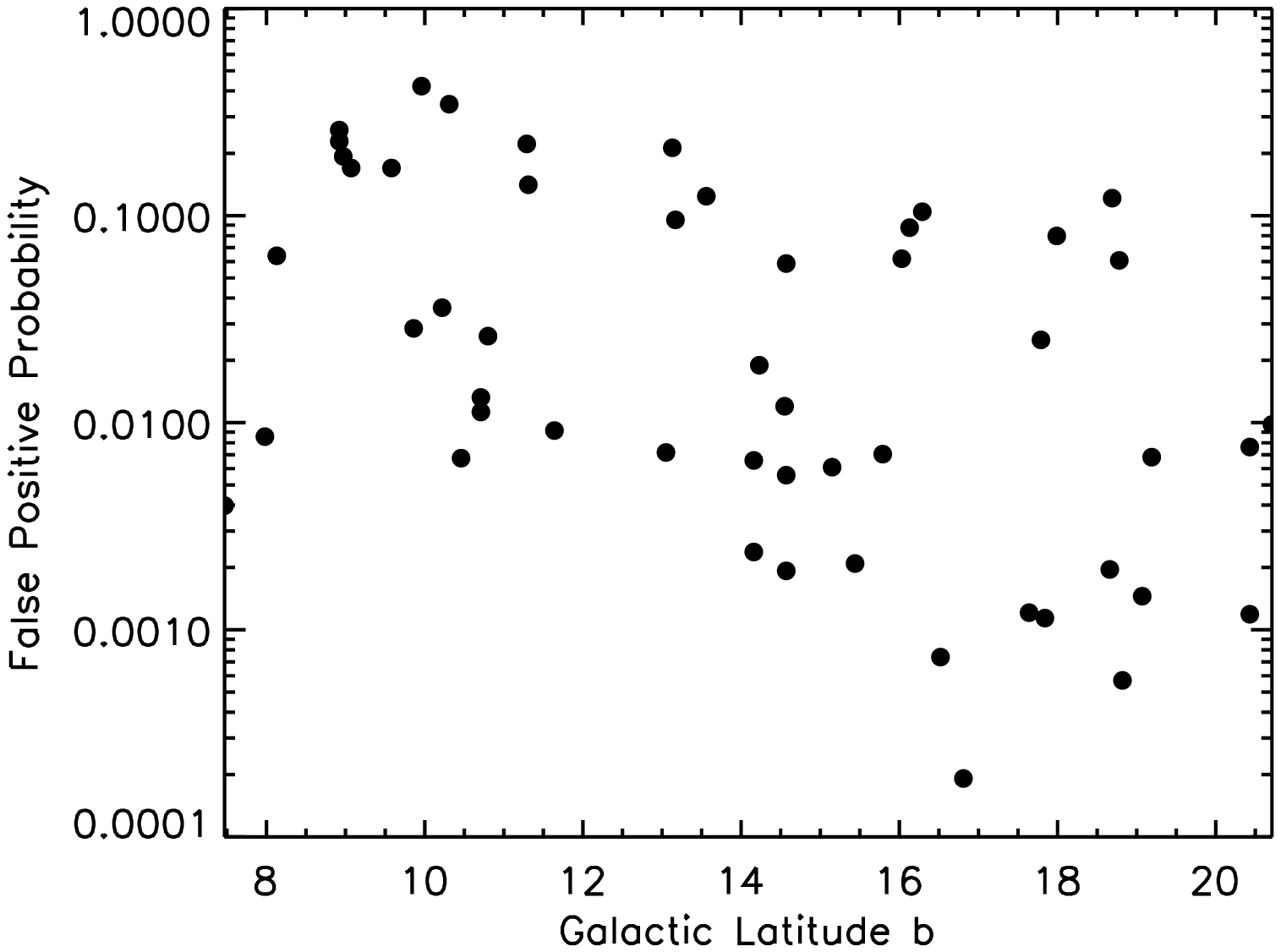}
\includegraphics[width=6in]{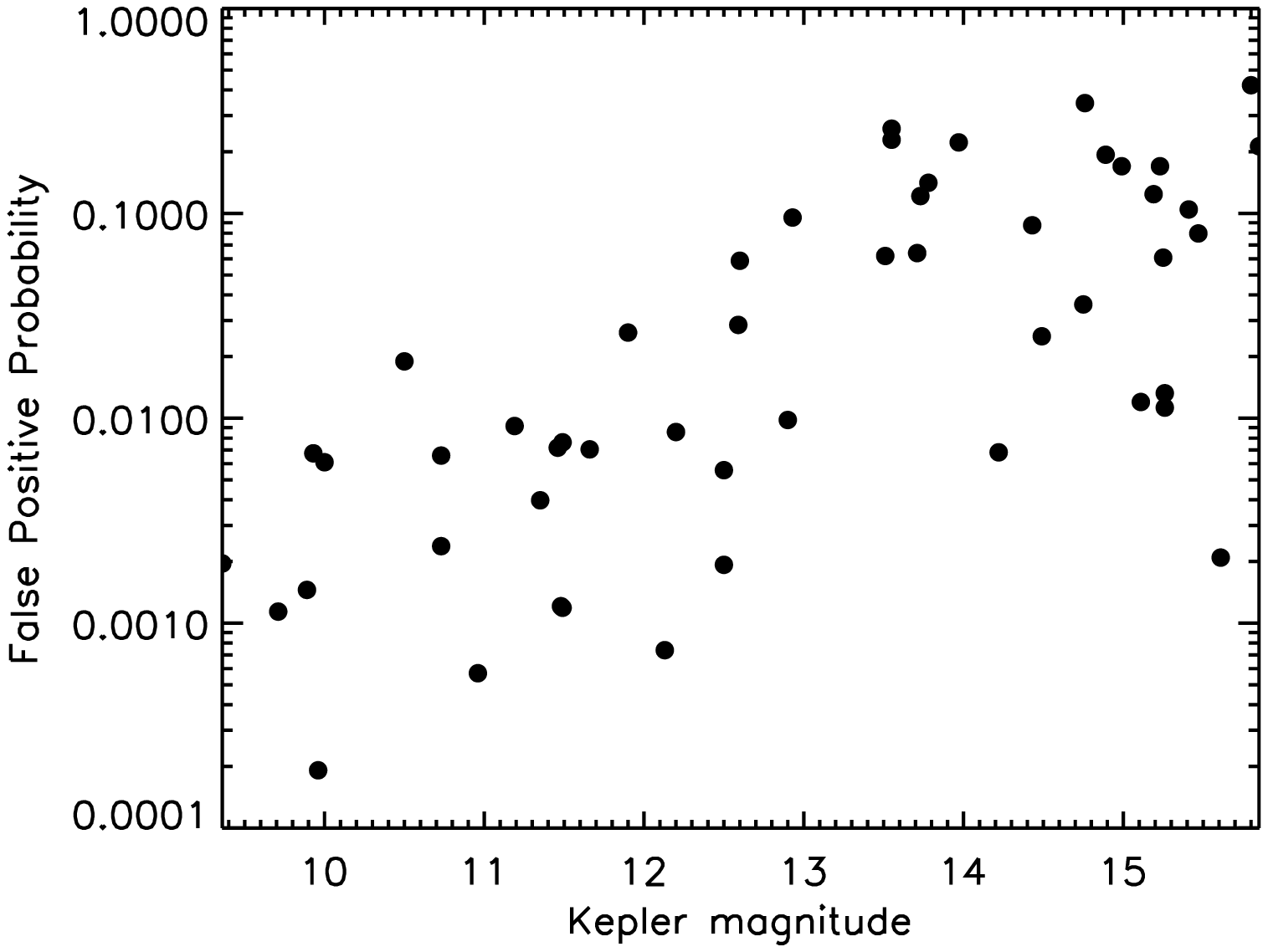}
 \caption{{\bf Top:} False Positive Probability (FPP) for each \kepler\ candidate (KOI) that we followed-up with \spitzer\ as a function of the target star's Galactic latitude. {\bf Bottom:} FPP as a function of the \kepler\ magnitude. The overall FPPs increase as the \kepler\ magnitude increases and the Galactic latitude decreases.}
  \label{fig:fppb}
\end{center}
\end{figure*}

We use two \spitzer\ Science Exploration Programs for this study.
We measure a median FPR of 0.8\% with a dispersion of 0.9\% from the KOIs of our first \spitzer\ program (60028). 
This value becomes 14\%, with a dispersion of 9\%, for the 15 targets from our second program (80117).
This is a higher FPR compared to the value derived from the first \spitzer\ program.
This is mainly because we do not have constraints from AO imaging nor from the \kepler\ centroid analysis for these KOIs, unlike for most of the targets from the first program.
This higher FPR value is also because the targeted sample is mainly composed of M-dwarf hosts. 
Low mass stars are usually faint in the \kepler\ bandpass (13.5 $<$mag$_{\mathrm{Kep}}<$ 16), hence the faint magnitude limit (Equation~\ref{eq:blender}) allows for fainter, and hence more numerous, EBs to mimic the transit signal than for the brighter targets in our first program.
This higher FPR value is also because the second sample contains mainly small size planet candidates (super-Earth candidates) which typically have smaller transit depths.
It is expected that the FPP increases with magnitude and with decreasing galactic latitude, and we observe such trends in figure~\ref{fig:fppb}). 
Furthermore, the faintness of M-stars also leads to a centroid analyzes with lower precisions, often prevents direct-imaging, and consequently contributes to a higher stellar blend frequency.

The stellar parameters that we use for this study are from \citet{Batalha12}. However, \citet{Huber13} recently updated the stellar properties using data from Quarter 1-16 and different observational techniques, but homogeneously extract the parameters from the Dartmouth stellar isochrones~\citep{Dotter08}. 
We find that the changes in stellar masses correspond to 9\% in average compared to~\citet{Batalha12} for the KOIs used for this study.
There are five KOIs of our list with updated masses that decrease between 15\% and 30\% (KOI-104, 244, 252, 663, 899, and 947). 
As noted by \citet{Huber13}, these are mainly low mass stars. 
Six stars have their updated masses that increase by more than 15\% (KOI-13, 87, 98, 111, 446, and 1362).
Since the mass of the primary is kept fixed in our study, we look at two extreme cases of mass changes (KOI-252 and 87) in order to test the effect of these variations on our determination of the FPPs.
\citet{Huber13} reported a mass for KOI-252 of 0.5~M$_{\odot}$, a decrease of 30\%, and a mass for KOI-87 that is 20\% larger (0.98~M$_{\odot}$).
For these two objects, we compute and compare the apparent transit depth ratios with the old and with the new masses for the primary stars, and we estimate the variations of these ratios for different masses of secondaries and tertiaries (such as presented in Fig~\ref{fig:ratiodepthd}).
We find that with a decrease in the stellar mass of 30\% for a 0.65~M$_{\odot}$ (KOI-252), the apparent transit ratios decrease of no more than 30\% for any secondaries with masses lower than 1.2~M$_{\odot}$, and distance modulii of 5.
Therefore, a primary star with a lower updated mass results in a lower number of low mass stars that can be rejected in order to satisfy the constraint from the {\spitzer} transit depth.
We compute the new FPP for KOI-252.01 considering the updated mass. We find that the FPP increases to 0.3\%, compared to 0.21\% with the previous mass.
Inversely, increasing the mass of the primary by 20\% (0.98~M$_{\odot}$, KOI-87) generates an increase in the apparent transit ratios of less than 30\% for any mass of the secondary that is lower than 1.2~M$_{\odot}$. The new FPP for this target goes down to 0.68\%.
We conclude that the new stellar masses computed by~\citet{Huber13} do not change significantly the FPPs of the KOIs that we have observed with {\spitzer}.
The FPPs presented in the current study are slightly underestimated for the KOIs for which the updated masses have decreased compared to~\citet{Batalha12}, whereas they are slightly overestimated for the KOIs for which the reevaluated masses are increased.

Finally, the period distribution of the \spitzer\ sample is skewed towards long period candidates compared to the distribution obtained from the list of KOIs (see Fig.~\ref{fig:period_distribution}).
This is because the second \spitzer\ sample was selected to focus on the super-Earth candidates that orbit in or close to the habitable zone of their host stars.
Longer period KOIs are more difficult to characterize (less \kepler\ transit events), therefore the skewed period distribution would tend to overestimate the FPR extrapolated from the \spitzer\ measurements.

\subsection{Comparison with complementary studies}
\label{sec:complementstudies}

The majority of the KOIs presented here have already been the focus of more specific studies, and have been validated or confirmed as true planets (see Table~\ref{tab:fpp}).
A subset of planets have constraints on their masses either from radial velocity measurements or from TTVs measured from the \kepler\ transit lightcurves \citep{Holman05,Agol05,Holman10}. 
Furthermore, nearly half (22) of the KOIs that we selected are in multiple transiting systems.
Eleven of these systems have been validated by \citet{Rowe14} using statistical arguments from \citet{Lissauer12, Lissauer14} and Q1-Q8 \kepler\ data.
Most of these systems were initially selected from the multiples in order to understand this relatively unexplored class of objects.
Others have companions that were discovered after their selection. 
Since then, we now know that candidates in multiple systems have a very high probability to be planets as demonstrated by~\citet{Lissauer11,Lissauer12,Lissauer14}.
However, for the purposes of this study, we assume that we have no information other than the depths and magnitudes measured in the \kepler\ and \spitzer\ bandpasses, and the direct imaging observations when available.

In many cases the \kepler\ team relies on the BLENDER procedure to assess the planetary nature of candidates in a statistical manner (e.g., Kepler-9d: \citealt{Torres11}, Kepler-11g: \citealt{Lissauer11}, and Kepler-10c: \citealt{Fressin11}). 
BLENDER takes into account the detailed morphology of the transit to reject as many false positive scenarios as possible. 
This approach is also used by other groups (e.g., \citealt{Nefs12,Diaz14}).
A candidate is considered validated when the likelihood of the signal being due to a true planet is much larger (by orders of magnitude) than the likelihood of a false positive. 
Many of the steps followed in the present analysis are inspired by the BLENDER approach, but are simplified and adapted to our purposes. 
In particular, we use here only the transit depth rather than its detailed shape to rule out blends.
In contrast to BLENDER, we adopt a single representative isochrone for all stars rather than different isochrones for the background binary and the target based on the measured spectroscopic properties. 
This latter approximation has little impact on our results. 
We test this for a sub-sample of our candidates for which we obtained spectroscopic reconnaissance of the main targets.
The sub-sample comprises some candidates that were used for BLENDER validation and have already been published (see Table~\ref{tab:fpp}).
For this subset, we determine the mass, radius, and age of the host star from a fit of the isochrones as described by \cite{Torres08}.
We compute a new FPP for each KOI in this subset. 
We compare this value to the FPP computed using the standard isochrone for the primary, and have checked that they are very similar.
This is because most of the false positive scenarios are mainly ruled-out from the combination of the \kepler\ photometry, the imaging and centroid information.
We also test the robustness of our method using isochrones of different ages for the secondary star instead of a standard isochrone.
The main motivation for this test is that a blended unassociated triple might have a component that is at a very different age from the primary.
We let the secondary isochrones range over all ages between 0.7 to 10 Gyr, while fixing the standard isochrone for the primary.
We find that the differences in the apparent transit depth ratios between \spitzer\ and \kepler\ (cf. Figure~\ref{fig:ratiodepthd}) vary from 10 to 30\% (in absolute) compared to the use of the standard isochrone for the secondary.
This difference depends on the mass of the secondary and tertiary, and it is partially degenerate with distance, which is a free parameter here.
In practice, this changes the blend frequency, and hence the FPP calculated in Section~\ref{sec:FPP} by only a small fraction.
The impact on the choice of isochrone ages for the secondary has also been tested by~\cite{Torres11} through the detailed study of the shape of transit light-curves using the BLENDER framework.
They find that the age of the secondary does not change significantly their estimate of the FPP. 
Therefore, we conclude that using generic isochrones does not affect the overall conclusion of this paper that the FPR of our sample is low.

\citet{Fressin13} perform detailed numerical simulations of the \kepler\ targets to predict the occurrence of astrophysical false positives, and its dependence on spectral type, candidate planet size, and orbital period.  
They find that the global false positive rate of \kepler\ is 9.4\%, peaking for giant planets (6--22\,$R_{\earth}$) at 17.7\%, reaching a low of 6.7\% for small Neptunes (2--4\,$R_{\earth}$), and increasing again for Earth-size planets (0.8--1.25\,$R_{\earth}$) to 12.3\%.
We compare these findings with the sample of candidates observed with \spitzer\ that fall in a similar overall size range (1.25--22\,$R_{\earth}$).
From the FPR estimated by \citet{Fressin13}, we can conclude that it would be extremely unlikely that we find no false positives in a random sample of 51 KOIs observed with \spitzer. 
The difference between our findings and the results from~\citet{Fressin13} is explained by the fact that our \spitzer\ sample underwent a much more stringent vetting procedure than typical KOIs.
Indeed, the estimated FPP for our sample is lower than 8.8\% at $3~\sigma$ level.

\begin{figure*}[h!]
\begin{center}
\includegraphics{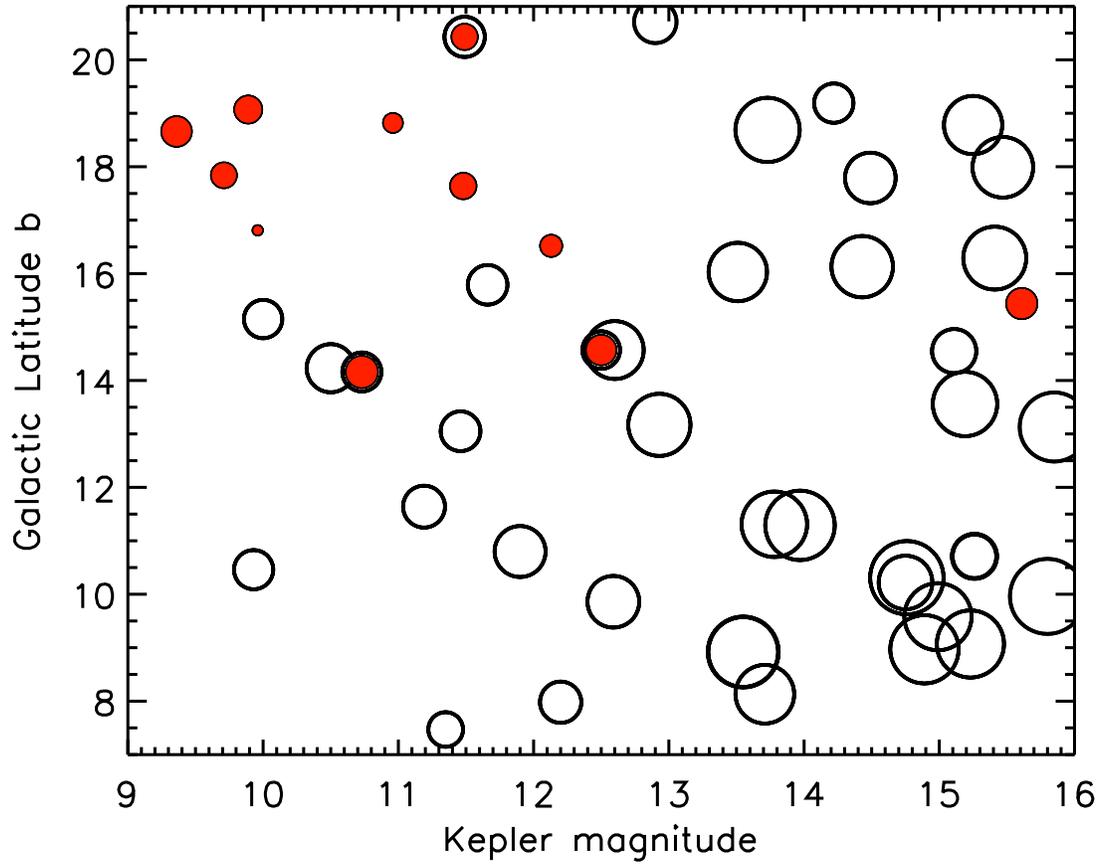}
 \caption{False Positive Probability (FPP) for each \kepler\ candidate (KOI) that we followed-up with \spitzer\ as a function of the target star's Galactic latitude and \kepler\ magnitude. The radii of the circles increase linearly as a function of the FPP value (largest circles for the largest FPP). The filled circle in red colors correspond to the targets for which the FPP is lower than 0.3\%, which could be considered as validated at $3~\sigma$ level of confidence. These validated planets represent one fourth of the overall sample. The FPP clearly increases towards the bottom-right corner as expected.}
  \label{fig:fppkpb}
\end{center}
\end{figure*}

We now compare our findings to the study of~\cite{Morton11}. 
These authors use the depths from the first 1235 KOIs reported from~\cite{Borucki11} together with generic assumptions and with the stellar population synthesis model TRILEGAL~\citep{Girardi05} to derive the FPP of all KOIs.
Their result was updated in \cite{Morton12} using the transit depths reported by~\cite{Batalha12}, and is based on 16 months of \kepler\ observations. 
The main steps employed in the current paper are similar to their approach. 
One of the main differences is that we include color information on the transits depths, thanks to the \spitzer\ observations (depths and magnitudes).
We also include constraints from direct imaging and centroids analyses when available. 
Another difference is that our work focuses on a small number of KOIs, whereas~\cite{Morton11, Morton12} aim at evaluating the FPR of the complete sample of KOIs.
Despite these differences, we find that our study is in good agreement with the study from~\cite{Morton11}.
They predict that the FPR would be lower than 5\% for half of the KOIs, and lower than 10\% for most of them.
Their results are consistent with the value that we report from our independent observational survey.
Furthermore, we also find trends in agreement with their work for the FPP values as a function of Galactic latitude and \kepler\ magnitude (see Figure~\ref{fig:fppkpb}).

There are other observational projects that address the question of the FPR in \kepler.
The most comparable observational study to our project in terms of candidate sample size is the work of~\cite{Santerne12}.
They find a relatively high value for the FPR of 34.8\% $\pm$ 6.5\% for their sample of 33 KOIs.
However, there are several important differences between our approach and the one employed by~\cite{Santerne12}.
The first difference remains in the observational method that is used to constrain the FPP of individual targets.
Instead of using transit photometry, \cite{Santerne12} obtain radial velocity observations to establish the nature of the transiting candidates.
The second important difference concerns the selection of the candidates considered for follow-up observations.
They focus on the deepest short-period transit signals with high $SNR$  that \kepler\ has detected.
Unlike for our study, \cite{Santerne12} restrict their targets to candidates with large transit depths that are greater than 0.4\%, and with short orbital periods lower than 25 days, and with host stars brighter than \kepler\ magnitude 14.7.
This selection is obviously driven by instrumental capabilities.
Instead, we select our candidates from a wider range of candidate sizes, orbital periods, and magnitudes that had been vetted by the {\it CFOP}.
Moreover, some of the KOIs selected by~\cite{Santerne12} were noted as being slightly V-shaped from the \kepler\ photometry in~\cite{Batalha12}, and these signals are considered as most likely due to EBs.
We emphasize that the FPR is expected to be greater for larger transit depths. 
Therefore, one should expect a higher FPR for the family of giant planets.

The high FPR found in the sample of~\cite{Santerne12} is consistent with the findings of~\cite{Demory11} for which close-in candidates were also targeted. 
\cite{Demory11} refine the photometric transit light curve analysis of 115 Kepler giant planet candidates based on photometric data from quarters Q0 to Q2.
These authors find that 14\% of these candidates are likely false positives based on the detections of their secondary eclipses.

Ground-based telescopes are also employed to examine the status of false-positives of a few KOIs using the same technique that we present here, i.e. color photometry.
\cite{Colon12} use the {\it GTC} telescope and observe transits of four short-period (P$<$6 days) planet candidates.

However, we note that the color photometric approach with ground-based instruments is limited to short period candidates for which the transit can be observed during the course of the night. 
Furthermore, the amplitude of the color-dependent effects for false positive detection increases as the two bandpasses under consideration are further removed in wavelength. 
Therefore, \spitzer\ is better adapted compared to ground-based photometric false-positive searches.

\section{Conclusion}
\label{sec:conclusion}

We present the results from two large observational campaigns, which were conducted with the \spitzer\ Space Telescope, dedicated to estimating the {\it false positive rate} amongst a sample of \kepler\ candidates.
We select a sub-sample of 51 candidates, spanning wide ranges in stellar, orbital and planetary parameter space, and we observe their transits with \spitzer\ at 4.5~\micron.
We measure the transit depths of these candidates in the \spitzer\ bandpass and compare them to the depths measured with \kepler.
This technique allows us to derive the probability that a false-positive (blended eclipsing binaries) could mimic the transit-shape signal. 
We estimate that 85\% of the KOIs from this sample have measured \kepler\ and \spitzer\ depths which agree at better than $3~\sigma$ level. 
We use the \spitzer\ observations to remove most of the red-dwarfs that could potentially remain as false positives.
By combining \spitzer\ and follow-up observations, we estimate that the overall false positive rate of our sample is estimated to be 1.3\%, and lower than 8.8\% at 99.7\% of confidence.
This rate implies that the vetting procedures of the \kepler\ data likely rule out a larger fraction of blends.
Extrapolating the empirical knowledge gained from this small sample to the overall \kepler\ sample of candidates, we find that the overall false positive rate of the \kepler\ sample is small. In this context, at least 90\% of the \kepler\ signals could be of planetary origin.

\acknowledgments

We thank the anonymous reviewer for the careful reading of our manuscript and the valuable comments.
This work is based on observations made with \kepler, which was competitively selected as the tenth Discovery mission. Funding for this mission is provided by NASA's Science Mission Directorate. The authors would like to thank the many people who generously gave so much their time to make this mission a success. 
This work is also based on observations made with the \spitzer\ {\it Space Telescope},
which is operated by the Jet Propulsion Laboratory, California Institute
of Technology under a contract with NASA. Support for this work was provided by NASA through an award issued by JPL/Caltech.
DC acknowledges support for this work from grants NNX09AB53G and NNX12AC77G, and GT acknowledges support from grant NNX12AC75G and NNX14AB83G, each from the NASA Kepler Mission Participating Scientist Program.
DC acknowledges the support of a grant from the John Templeton Foundation. The opinions expressed in this publication are those of the authors and do not necessarily reflect the views of the John Templeton Foundation.
We would like to thank the Spitzer staff at IPAC and in particular Nancy Silbermann for scheduling the Spitzer observations of this program.
J.-M.D. and S.B. acknowledge the Sagan Exoplanet Fellowship program supported by the National Aeronautics and Space Administration and administered by the NASA Exoplanet Science Institute (NExScI).
We thank Samaya Nissanke for careful reading of the manuscript.


\begin{figure*}[h!]
\begin{center}
\includegraphics{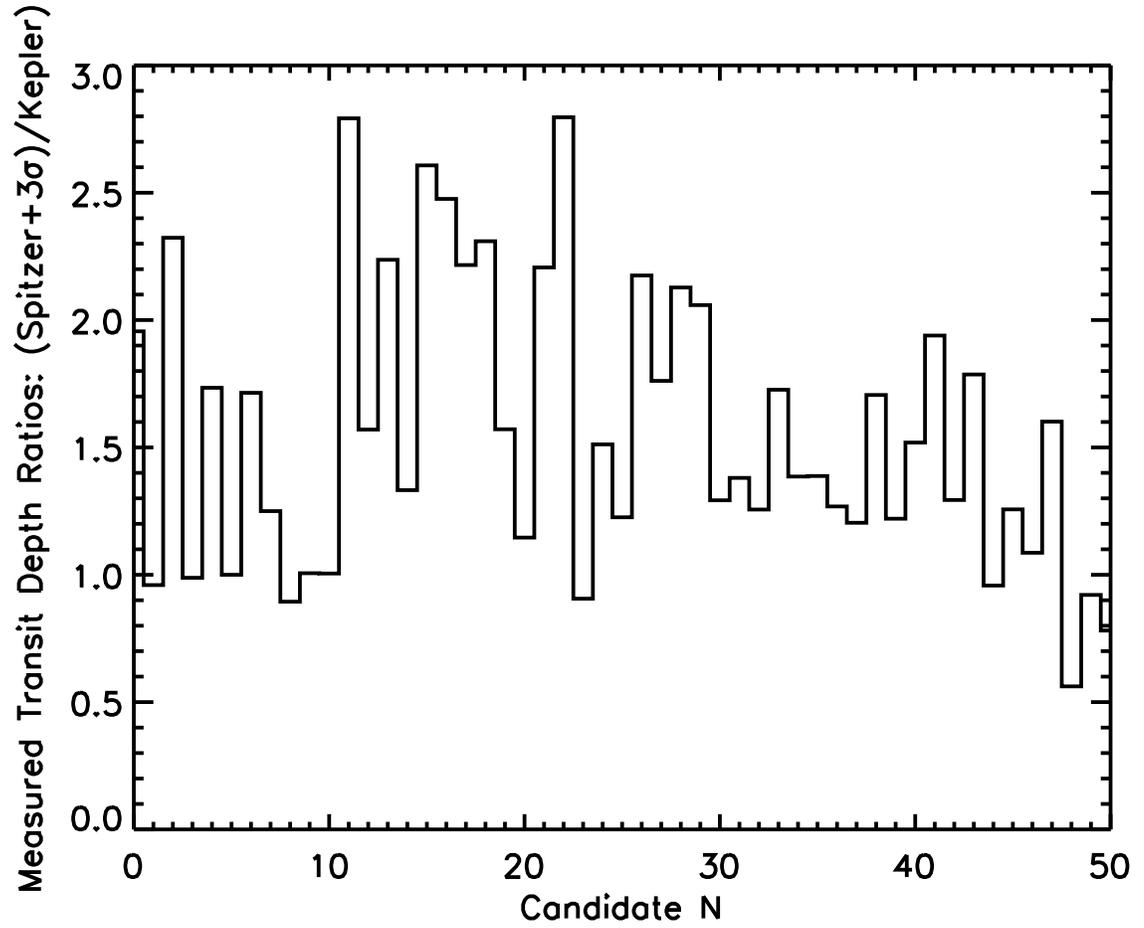}
 \caption{Ratios of the transit depths measured from the \spitzer\ 4.5~\micron\ lightcurves ($+3~\sigma$) to the transit depths measured in the \kepler\ bandpass for 50 KOIs targeted in this program. The KOIs are ordered similarly to Figure~\ref{fig:depth}: increasing transit depths from \kepler\ towards the right.}
  \label{fig:fpp}
\end{center}
\end{figure*}

\begin{figure*}[h!]
\begin{center}
\includegraphics{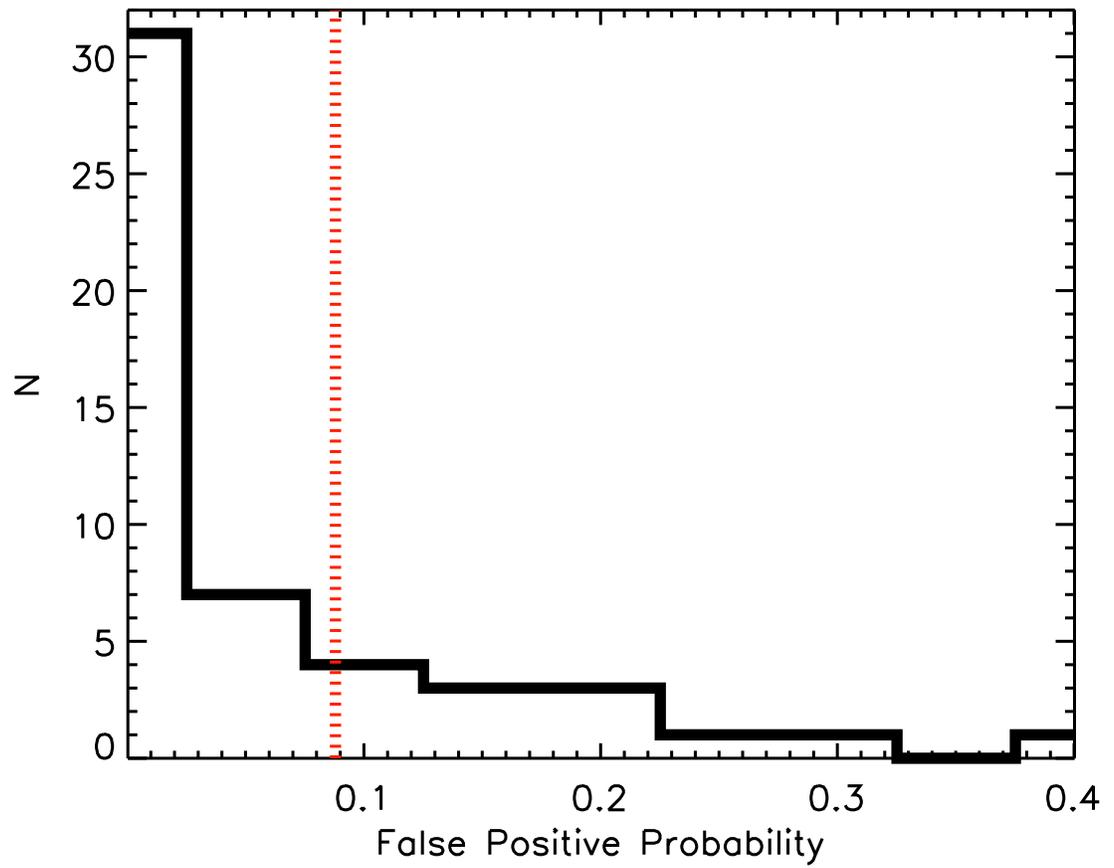}
 \caption{Histogram distribution of the False Positive Probability (FPP) of the \kepler\ candidates (KOIs) that we observed with \spitzer. Half of the overall sample has a FPP$<1\%$. The vertical dashed line shows the 3$\sigma$ upper limit of the FPR (8.8\%) of the KOIs we present in this project (see Section~\ref{sec:onfpp}).}
  \label{fig:fpp}
\end{center}
\end{figure*}

\newpage


\newpage

\begin{center}
\begin{deluxetable*}{cccccc}
\tablewidth{0pc}
\tablecaption{Measurements from \spitzer\ observations of program ID 60028.
\label{tab:spitzer2}}
\tablehead{
\colhead{KOI}                             &
\colhead{AOR}                            &
\colhead{Magnitude}                           &
\colhead{Flux}                            &
\colhead{Depth$_{Kepler}$}            &
\colhead{Depth$_{Spitzer}$}            \\
\colhead{}                        &
\colhead{}                      &
\colhead{}                         &
\colhead{(mJy)}                         &
\colhead{(\%)}                         &
\colhead{(\%)}                         
}
\startdata
12.01 & r39525120 & $10.183\pm0.005$ & $15.179\pm0.168$ & $0.9271\pm0.0019$ & $0.688^{+0.012}_{-0.011}$   \\
13.01 & r39525376 & $9.397\pm0.002$ & $31.305\pm0.115$ & $0.4646\pm0.0031$ & $0.228^{+0.011}_{-0.011}$  \\
42.01 & r41010688 & $8.084\pm0.003$ & $104.939\pm0.753$ & $0.0334\pm0.0004$ & $0.018^{+0.005}_{-0.005}$  \\
69.01 & r41009920 & $8.331\pm0.003$ & $83.561\pm0.640$ & $0.0271\pm0.0003$ & $0.007^{+0.008}_{-0.006}$  \\
69.01 & r41010432 & $8.332\pm0.004$ & $83.489\pm0.768$ & $0.0271\pm0.0003$ & $0.009^{+0.008}_{-0.007}$  \\
70.01 & r41165824 & $10.855\pm0.006$ & $8.176\pm0.119$ & $0.1028\pm0.0021$ & $0.057^{+0.022}_{-0.023}$  \\
70.03 & r39437568 & $10.834\pm0.012$ & $8.337\pm0.221$ & $0.0829\pm0.0019$ & $0.055^{+0.016}_{-0.017}$  \\
70.03 & r41164544 & $10.841\pm0.015$ & $8.283\pm0.289$ & $0.0829\pm0.0019$ & $0.062^{+0.016}_{-0.016}$  \\
72.02 & r39369984 & $9.462\pm0.007$ & $29.484\pm0.449$ & $0.0497\pm0.0005$ & $0.032^{+0.008}_{-0.008}$  \\
72.02 & r39369216 & $9.476\pm0.016$ & $29.115\pm1.076$ & $0.0497\pm0.0005$ & $0.033^{+0.010}_{-0.010}$  \\
82.01 & r39420672 & $9.353\pm0.003$ & $32.605\pm0.258$ & $0.0949\pm0.0034$ & $0.056^{+0.010}_{-0.010}$  \\
82.02 & r39419904 & $9.348\pm0.003$ & $32.748\pm0.207$ & $0.0271\pm0.0011$ & $0.038^{+0.011}_{-0.011}$  \\
82.02 & r39437056 & $9.353\pm0.004$ & $32.617\pm0.283$ & $0.0271\pm0.0011$ & $0.018^{+0.012}_{-0.011}$  \\
84.01 & r39369472 & $10.295\pm0.005$ & $13.689\pm0.167$ & $0.0698\pm0.0016$ & $0.110^{+0.020}_{-0.022}$  \\
84.01 & r39370240 & $10.305\pm0.003$ & $13.573\pm0.100$ & $0.0698\pm0.0016$ & $0.016^{+0.017}_{-0.013}$  \\
84.01 & r41165312 & $10.312\pm0.004$ & $13.481\pm0.123$ & $0.0698\pm0.0016$ & $0.070^{+0.019}_{-0.018}$  \\
94.01 & r39421440 & $10.954\pm0.012$ & $7.463\pm0.210$ & $0.5745\pm0.0016$ & $0.478^{+0.017}_{-0.017}$  \\
98.01 & r39421184 & $10.991\pm0.004$ & $7.213\pm0.072$ & $0.2276\pm0.0006$ & $0.170^{+0.016}_{-0.016}$  \\
103.01 & r39366400 & $11.032\pm0.006$ & $6.944\pm0.091$ & $0.0821\pm0.0114$ & $0.069^{+0.028}_{-0.028}$  \\
103.01 & r39366144 & $11.013\pm0.012$ & $7.067\pm0.192$ & $0.0821\pm0.0114$ & $0.075^{+0.025}_{-0.026}$  \\
104.01 & r39420160 & $10.601\pm0.004$ & $10.327\pm0.096$ & $0.1501\pm0.0219$ & $0.081^{+0.039}_{-0.043}$  \\
104.01 & r41163776 & $10.586\pm0.004$ & $10.474\pm0.103$ & $0.1501\pm0.0219$ & $0.154^{+0.039}_{-0.039}$  \\
137.01 & r39365888 & $11.755\pm0.014$ & $3.567\pm0.116$ & $0.2292\pm0.0054$ & $0.168^{+0.037}_{-0.040}$ \\
137.02 & r39369728 & $11.901\pm0.023$ & $3.119\pm0.164$ & $0.3235\pm0.0087$ & $0.371^{+0.046}_{-0.049}$ \\
157.03 & r41197568 & $12.196\pm0.022$ & $2.378\pm0.117$ & $0.1401\pm0.0047$ & $0.046^{+0.040}_{-0.035}$  \\
157.03 & r41197312 & $12.194\pm0.017$ & $2.381\pm0.091$ & $0.1401\pm0.0047$ & $0.138^{+0.040}_{-0.039}$ \\
244.01 & r39437312 & $9.519\pm0.003$ & $27.997\pm0.221$ & $0.1188\pm0.0012$ & $0.125^{+0.012}_{-0.013}$ \\
244.02 & r39438848 & $9.494\pm0.006$ & $28.630\pm0.422$ & $0.0400\pm0.0003$ & $0.041^{+0.011}_{-0.011}$ \\
244.02 & r39439104 & $9.483\pm0.002$ & $28.926\pm0.137$ & $0.0400\pm0.0003$ & $0.011^{+0.011}_{-0.009}$ \\
244.02 & r41165568 & $9.497\pm0.005$ & $28.547\pm0.310$ & $0.0400\pm0.0003$ & $0.043^{+0.011}_{-0.012}$ \\
245.01 & r39420928 & $8.084\pm0.010$ & $104.914\pm2.426$ & $0.0607\pm0.0017$ & $0.051^{+0.009}_{-0.009}$ \\
245.01 & r41009664 & $7.865\pm0.002$ & $128.432\pm0.526$ & $0.0607\pm0.0017$ & $0.049^{+0.005}_{-0.005}$ \\
246.01 & r41009408 & $8.538\pm0.005$ & $69.088\pm0.773$ & $0.0350\pm0.0003$ & $0.056^{+0.007}_{-0.007}$ \\
246.01 & r41010176 & $8.533\pm0.005$ & $69.385\pm0.769$ & $0.0350\pm0.0003$ & $0.035^{+0.007}_{-0.007}$ \\
247.01 & r39368704 & $11.017\pm0.010$ & $7.042\pm0.165$ & $0.0992\pm0.0241$ & $0.056^{+0.038}_{-0.037}$ \\
247.01 & r39368448 & $11.013\pm0.022$ & $7.071\pm0.350$ & $0.0992\pm0.0241$ & $0.105^{+0.034}_{-0.035}$ \\
247.01 & r41164032 & $11.034\pm0.006$ & $6.935\pm0.096$ & $0.0992\pm0.0241$ & $0.089^{+0.048}_{-0.046}$ \\
248.01 & r39370496 & $12.276\pm0.128$ & $2.210\pm0.564$ & $0.1762\pm0.0187$ & $0.078^{+0.052}_{-0.051}$ \\
248.01 & r41165056 & $12.280\pm0.027$ & $2.200\pm0.132$ & $0.1762\pm0.0187$ & $0.135^{+0.049}_{-0.050}$ \\
248.02 & r39366912 & $12.298\pm0.024$ & $2.165\pm0.115$ & $0.1048\pm0.0182$ & $0.086^{+0.057}_{-0.056}$ \\
248.02 & r39367168 & $12.278\pm0.015$ & $2.204\pm0.075$ & $0.1048\pm0.0182$ & $0.087^{+0.061}_{-0.059}$ \\
248.02 & r39366656 & $12.269\pm0.022$ & $2.224\pm0.109$ & $0.1048\pm0.0182$ & $0.218^{+0.071}_{-0.077}$ \\
249.01 & r39419648 & $11.016\pm0.005$ & $7.052\pm0.074$ & $0.1640\pm0.0020$ & $0.174^{+0.032}_{-0.033}$ \\
249.01 & r39421952 & $11.016\pm0.005$ & $7.052\pm0.074$ & $0.1640\pm0.0020$ & $0.106^{+0.030}_{-0.032}$ \\
250.02 & r41197056 & $12.521\pm0.039$ & $1.763\pm0.153$ & $0.1896\pm0.0103$ & $0.108^{+0.066}_{-0.069}$ \\
250.02 & r41196800 & $12.513\pm0.040$ & $1.776\pm0.158$ & $0.1896\pm0.0103$ & $0.280^{+0.063}_{-0.067}$ \\
250.02 & r41196544 & $12.530\pm0.028$ & $1.749\pm0.107$ & $0.1896\pm0.0103$ & $0.105^{+0.068}_{-0.071}$ \\
251.01 & r39437824 & $11.485\pm0.012$ & $4.578\pm0.127$ & $0.2228\pm0.0425$ & $0.280^{+0.041}_{-0.041}$ \\
251.01 & r41164800 & $11.461\pm0.010$ & $4.681\pm0.110$ & $0.2228\pm0.0425$ & $0.315^{+0.075}_{-0.083}$ \\
252.01 & r39421696 & $12.489\pm0.029$ & $1.815\pm0.117$ & $0.2157\pm0.0726$ & $0.152^{+0.051}_{-0.052}$ \\
252.01 & r41166336 & $12.488\pm0.023$ & $1.818\pm0.093$ & $0.2157\pm0.0726$ & $0.187^{+0.047}_{-0.049}$ \\
253.01 & r41440256 & $12.318\pm0.010$ & $2.125\pm0.048$ & $0.1747\pm0.1242$ & $0.094^{+0.066}_{-0.068}$ \\
255.01 & r39420416 & $11.998\pm0.015$ & $2.853\pm0.094$ & $0.2393\pm0.0636$ & $0.143^{+0.039}_{-0.039}$ \\
260.02 & r39438080 & $9.320\pm0.004$ & $33.626\pm0.304$ & $0.0346\pm0.0006$ & $0.039^{+0.007}_{-0.007}$ \\
271.01 & r39439360 & $10.236\pm0.007$ & $14.464\pm0.241$ & $0.0350\pm0.0008$ & $0.013^{+0.015}_{-0.011}$  \\
271.01 & r41166080 & $10.236\pm0.009$ & $14.463\pm0.293$ & $0.0350\pm0.0008$ & $0.005^{+0.011}_{-0.005}$  \\
273.01 & r39368192 & $9.953\pm0.002$ & $18.763\pm0.078$ & $0.0297\pm0.0101$ & $0.029^{+0.024}_{-0.022}$  \\
273.01 & r39367680 & $9.965\pm0.012$ & $18.558\pm0.496$ & $0.0297\pm0.0101$ & $0.041^{+0.020}_{-0.022}$  \\
273.01 & r39367424 & $9.959\pm0.004$ & $18.661\pm0.169$ & $0.0297\pm0.0101$ & $0.029^{+0.019}_{-0.020}$  \\
314.01 & r44144384 & $9.322\pm0.003$ & $33.568\pm0.234$ & $0.0740\pm0.0139$ & $0.101^{+0.020}_{-0.021}$  \\
365.01 & r40252928 & $9.620\pm0.004$ & $25.504\pm0.211$ & $0.0656\pm0.0039$ & $0.086^{+0.010}_{-0.011}$  \\
365.01 & r40252672 & $9.616\pm0.009$ & $25.603\pm0.528$ & $0.0656\pm0.0039$ & $0.078^{+0.010}_{-0.010}$ \\
\enddata
\label{tab:spitzer1}
\end{deluxetable*}
\end{center}

\begin{center}
\begin{deluxetable*}{cccccc}
\tablewidth{0pc}
\tablecaption{Measurements from \spitzer\ observations of program ID 80117.
\label{tab:spitzer2}}
\tablehead{
\colhead{KOI}                             &
\colhead{AOR}                            &
\colhead{Magnitude}                           &
\colhead{Flux}                            &
\colhead{Depth$_{Kepler}$}            &
\colhead{Depth$_{Spitzer}$}           \\
\colhead{}                        &
\colhead{}                      &
\colhead{}                         &
\colhead{(mJy)}                         &
\colhead{(\%)}                         &
\colhead{(\%)}                         
}
\startdata
87.01 & r44159488 & $10.128\pm0.004$ & $15.965\pm0.163$ & $0.0492\pm0.0075$ & $0.008^{+0.012}_{-0.007}$  \\
111.03 & r44162048 & $11.206\pm0.017$ & $5.916\pm0.229$ & $0.0616\pm0.0012$ & $0.094^{+0.025}_{-0.026}$  \\
174.01 & r44162560 & $11.515\pm0.011$ & $4.453\pm0.115$ & $0.1039\pm0.0026$ & $0.120^{+0.036}_{-0.038}$  \\
174.01 & r44162304 & $11.508\pm0.024$ & $4.483\pm0.238$ & $0.1039\pm0.0026$ & $0.098^{+0.032}_{-0.034}$  \\
446.02 & r44161024 & $12.042\pm0.023$ & $2.739\pm0.142$ & $0.0920\pm0.0365$ & $0.213^{+0.060}_{-0.059}$  \\
446.02 & r44161536 & $12.033\pm0.046$ & $2.764\pm0.278$ & $0.0920\pm0.0365$ & $0.118^{+0.051}_{-0.054}$  \\
446.02 & r44160768 & $12.047\pm0.028$ & $2.727\pm0.170$ & $0.0920\pm0.0365$ & $0.032^{+0.050}_{-0.030}$  \\
555.02 & r44162816 & $12.969\pm0.136$ & $1.166\pm0.313$ & $0.0937\pm0.0030$ & $0.103^{+0.051}_{-0.053}$  \\
663.02 & r44159232 & $10.751\pm0.014$ & $9.002\pm0.286$ & $0.0693\pm0.0083$ & $0.104^{+0.026}_{-0.026}$  \\
663.02 & r44158720 & $10.765\pm0.007$ & $8.882\pm0.142$ & $0.0693\pm0.0083$ & $0.083^{+0.029}_{-0.030}$  \\
701.03 & r44163840 & $11.630\pm0.016$ & $4.003\pm0.145$ & $0.0719\pm0.0108$ & $0.100^{+0.026}_{-0.026}$  \\
711.03 & r44158976 & $12.313\pm0.098$ & $2.135\pm0.430$ & $0.0698\pm0.0029$ & $0.068^{+0.037}_{-0.038}$  \\
817.01 & r44160512 & $12.216\pm0.053$ & $2.334\pm0.268$ & $0.1122\pm0.0576$ & $0.058^{+0.080}_{-0.051}$  \\
817.01 & r44160256 & $12.211\pm0.027$ & $2.344\pm0.140$ & $0.1122\pm0.0576$ & $0.073^{+0.076}_{-0.062}$  \\
854.01 & r44164864 & $12.363\pm0.027$ & $2.039\pm0.125$ & $0.1694\pm0.1658$ & $0.043^{+0.048}_{-0.035}$  \\
854.01 & r44164352 & $12.363\pm0.117$ & $2.039\pm0.483$ & $0.1694\pm0.1658$ & $0.162^{+0.048}_{-0.052}$  \\
899.03 & r44165376 & $11.743\pm0.011$ & $3.609\pm0.092$ & $0.0762\pm0.0214$ & $0.057^{+0.041}_{-0.040}$  \\
899.03 & r44165632 & $11.736\pm0.021$ & $3.633\pm0.171$ & $0.0762\pm0.0214$ & $0.137^{+0.044}_{-0.045}$  \\
899.03 & r44166144 & $11.748\pm0.020$ & $3.594\pm0.160$ & $0.0762\pm0.0214$ & $0.122^{+0.046}_{-0.048}$  \\
947.01 & r44164608 & $11.889\pm0.025$ & $3.154\pm0.175$ & $0.1607\pm0.0177$ & $0.113^{+0.041}_{-0.047}$  \\
947.01 & r44165120 & $11.898\pm0.021$ & $3.128\pm0.146$ & $0.1607\pm0.0177$ & $0.124^{+0.051}_{-0.053}$  \\
952.03 & r44159744 & $12.604\pm0.037$ & $1.633\pm0.134$ & $0.1939\pm0.0543$ & $0.146^{+0.065}_{-0.066}$  \\
952.03 & r44160000 & $12.602\pm0.034$ & $1.635\pm0.123$ & $0.1939\pm0.0543$ & $0.324^{+0.065}_{-0.066}$  \\
1199.01 & r44166400 & $12.681\pm0.134$ & $1.520\pm0.403$ & $0.1039\pm0.0625$ & $0.085^{+0.044}_{-0.047}$  \\
1361.01 & r44161280 & $12.224\pm0.061$ & $2.316\pm0.302$ & $0.1419\pm0.0198$ & $0.104^{+0.044}_{-0.047}$  \\
\enddata
\label{tab:spitzer2}
\end{deluxetable*}
\end{center}

\begin{center}
\begin{deluxetable*}{ccccccccc}
\tablewidth{0pc}
\tablecaption{FPP results.
\label{tab:fpp}}
\tablehead{
\colhead{KOI}                             &
\colhead{\ensuremath{\sigma_{\rm K-S}}}  &
\colhead{Gal. long}                            &
\colhead{Gal. lat}                            &
\colhead{Magnitude}                           &
\colhead{AO}                            &
\colhead{Centroid}            &
\colhead{FPP}               &
\colhead{comments}             \\    
\colhead{}                        &
\colhead{(*)}                        &
\colhead{l (deg)}                      &
\colhead{b (deg)}                         &
\colhead{(Kepler)}                         &
\colhead{(**)}                         &
\colhead{(**)}                         &
\colhead{(\%)}                         &
\colhead{}
}
\startdata
  12.01 &  19.7 & 75.50 &  7.47 & 11.35 & n & y &  0.40 & \\
  13.01 &  20.7 & 77.51 & 16.81 &  9.96 & y & n &  0.02 &  Kepler-13b \cite{Barnes11}\\
  42.01 &   3.1 & 74.81 & 18.66 &  9.36 & y & n &  0.20 &   Kepler-410A b  \cite{VanEylen14}\\
  69.01 &   3.4 & 71.20 & 10.46 &  9.93 & y & n &  0.67 &   Kepler-93b \cite{Ballard14}\\
  70.01 &   2.0 & 73.38 & 14.57 & 12.50 & y & n &  0.19 &  Kepler-20b \cite{Gautier12}\\
  70.03 &   2.0 & 73.38 & 14.57 & 12.50 & y & n &  0.56 &  Kepler-20c \cite{Gautier12}\\
  72.02 &   2.8 & 80.49 & 18.82 & 10.96 & y & n &  0.06 &  Kepler-10c \cite{Fressin11}\\
  82.01 &   3.7 & 76.48 & 20.43 & 11.49 & y & n &  0.12 &  Kepler-102e \cite{Marcy14}\\
  82.02 &   0.2 & 76.48 & 20.43 & 11.49 & y & n &  0.76 &  Kepler-102d \cite{Marcy14}\\
  84.01 &   0.9 & 70.10 & 10.80 & 11.90 & y & n &  2.62 &  Kepler-19b \cite{Ballard11} \\
  94.01 &   5.7 & 76.23 &  7.98 & 12.20 & y & n &  0.86 &   Kepler-89d \cite{Weiss13} \\
  98.01 &   3.6 & 78.15 & 16.52 & 12.13 & y & n &  0.07 &  Kepler-14b \cite{Buchhave11} \\
 103.01 &   0.4 & 70.46 &  9.86 & 12.59 & y & n &  2.85 &  TTVs\\
 104.01 &   0.8 & 76.71 & 20.71 & 12.90 & y & n &  0.98 &  Kepler-94b \cite{Marcy14} \\
 137.01 &   1.5 & 79.01 &  8.92 & 13.55 & n & n & 25.88 &  Kepler-18c \cite{Cochran11} \\
 137.02 &   0.9 & 79.01 &  8.92 & 13.55 & n & n & 22.86 &  Kepler-18d \cite{Cochran11} \\
 157.03 &   1.7 & 76.16 &  8.13 & 13.71 & n & y &  6.40 &  Kepler-11b \cite{Lissauer11} \\
 244.01 &   0.5 & 70.35 & 14.16 & 10.73 & y & n &  0.24 & Kepler-25c \cite{Steffen12} \\
 244.02 &   1.3 & 70.35 & 14.16 & 10.73 & y & n &  0.66 & Kepler-25b \cite{Steffen12} \\
 245.01 &   2.4 & 74.44 & 17.84 &  9.71 & y & n &  0.11 &  Kepler-37b \cite{Barclay13} \\
 246.01 &   2.1 & 80.69 & 15.15 & 10.00 & y & n &  0.61 &  Kepler-68b \cite{Gilliland13} \\
 247.01 &   0.5 & 80.24 & 19.19 & 14.22 & y & n &  0.68 & \\
 248.01 &   1.7 & 73.26 & 10.71 & 15.26 & n & y &  1.13 &  Kepler-49b \cite{Steffen13} \\
 248.02 &   0.3 & 73.26 & 10.71 & 15.26 & n & y &  1.32 &  Kepler-49c \cite{Steffen13} \\
 249.01 &   1.1 & 76.09 & 17.79 & 14.49 & y & n &  2.51 & \\
 250.02 &   0.5 & 76.68 & 17.99 & 15.47 & n & n &  7.98 &  Kepler-26c \cite{Steffen12} \\
 251.01 &   1.1 & 81.59 & 10.22 & 14.75 & y & n &  3.59 &  Kepler-125b \cite{Rowe14} \\
 252.01 &   0.6 & 80.31 & 15.44 & 15.61 & n & y &  0.21 & \\
 253.01 &   0.6 & 80.19 & 18.78 & 15.25 & n & n &  6.08 & \\
 255.01 &   1.3 & 73.62 & 14.55 & 15.11 & n & y &  1.20 & \\
 260.02 &   0.6 & 75.64 & 14.23 & 10.50 & y & n &  1.89 & Kepler-126d \cite{Rowe14} \\
 271.01 &   3.0 & 76.22 & 17.64 & 11.48 & y & n &  0.12 & Kepler-127d \cite{Rowe14} \\
 273.01 &   0.2 & 69.44 & 13.05 & 11.46 & y & n &  0.72 & \\
 314.01 &   1.1 & 75.11 & 13.17 & 12.93 & n & n &  9.54 & Kepler-138c \cite{Rowe14} \\
 365.01 &   1.9 & 83.14 & 11.64 & 11.19 & y & n &  0.92 & \\
      \\
\hline              
     \\
  87.01 &   2.9 & 79.09 & 15.79 & 11.66 & y & n &  0.70 &  Kepler-22b \cite{Borucki12} \\
 111.03 &   1.2 & 73.19 & 14.57 & 12.60 & n & n &  5.87 & Kepler-104d \cite{Rowe14} \\
 174.01 &   0.1 & 81.57 & 11.31 & 13.78 & n & n & 14.09 & \\
 446.02 &   0.4 & 69.82 & 16.13 & 14.43 & n & n &  8.74 & Kepler-157b \cite{Rowe14} \\
 555.02 &   0.2 & 73.86 & 10.31 & 14.76 & n & n & 34.51 & \\
 663.02 &   1.2 & 72.20 & 16.03 & 13.51 & y & n &  6.19 & Kepler-205c \cite{Rowe14} \\
 701.03 &   1.0 & 75.04 & 18.69 & 13.73 & n & n & 12.14 & Kepler-62e \cite{Borucki13} \\
 711.03 &   0.1 & 79.45 & 11.29 & 13.97 & n & n & 22.19 & \\
 817.01 &   0.6 & 69.88 & 16.29 & 15.41 & n & n & 10.44 & Kepler-236c \cite{Rowe14} \\
 854.01 &   0.4 & 73.45 & 13.13 & 15.85 & n & n & 21.25 & \\
 899.03 &   0.7 & 77.66 &  9.07 & 15.23 & n & n & 16.97 & Kepler-249d \cite{Rowe14} \\
 947.01 &   1.1 & 78.48 & 13.56 & 15.19 & n & n & 12.40 & \\
 952.03 &   0.6 & 80.54 &  9.96 & 15.80 & n & n & 42.18 & Kepler-32b \cite{Fabrycky12} \\
1199.01 &   0.2 & 72.28 &  8.97 & 14.89 & n & n & 19.32 & \\
1361.01 &   0.7 & 76.01 &  9.58 & 14.99 & y & n & 16.97 &  Kepler-61b \cite{Ballard13} \\

\enddata
\label{tab:fpp}
\tablenotetext{*}{Difference between the \spitzer\ and \kepler\ apparent transit depths in unit of $\sigma$ from combined measurements. The differences are not corrected for dilution caused by the presence of a close-by stellar companions.}
\tablenotetext{**}{These columns indicate whether Information from the Adaptive-Optic imaging (AO) followup and \kepler\ centroid analysis are available: "y" means that information is available and used in the study, "no" means that no information on AO and centroid have been used.}
\end{deluxetable*}
\end{center}


\end{document}